\documentclass[10pt,twoside,letter,amsmath,amssymb,aps,showkeys,showpacs, twocolumn]{revtex4}
\usepackage{amsfonts}
\usepackage{graphics}
\usepackage{graphicx}
\usepackage{epsfig}
\usepackage{color}
\usepackage[T1]{fontenc}

\begin{document}


\thispagestyle{myheadings}


\title{One-loop electroweak corrections for polarized M{\o}ller scattering\\
at different renormalization schemes and conditions}

\author{Aleksandrs Aleksejevs}
\email{aaleksejevs@swgc.mun.ca}
\affiliation{Memorial University, Corner Brook, Canada}

\author{Svetlana Barkanova}
\email{svetlana.barkanova@acadiau.ca}
\affiliation{Acadia University, Wolfville, Canada}

\author{Alexander Ilyichev}
\email{ily@hep.by}
\affiliation{
National Center of Particle and High Energy Physics, Minsk, Belarus}

\author{Yury Kolomensky}
\email{yury@physics.berkeley.edu}
\affiliation{University of California, Berkeley, USA}

\author{Vladimir Zykunov}
\email{vladimir.zykunov@cern.ch}
\affiliation{Belarussian State University of Transport, Gomel, Belarus}

\date{16/08/2011}

\begin{abstract}
Using two different approaches, we perform updated and
detailed calculations of the complete one-loop (Next-to-Leading Order (NLO)) set of electroweak radiative corrections to the parity violating 
$e^-e^- \rightarrow e^-e^- (\gamma)$ scattering asymmetry.
Our first approach, more classical, relies on calculations "by hand" with reasonable approximations. 
Our second approach relies on program packages FeynArts, FormCalc, LoopTools, and FORM. 
The detailed numerical analysis of the various
contributions is provided for a wide range of energies relevant for the ultra-precise 11 GeV MOLLER 
experiment planned at the Jefferson Laboratory, as well as future experiments at the International Linear Collider (ILC). 
The numerical results obtained within the on-shell renormalization scheme
using two different sets of renormalization conditions are in excellent agreement. 
We also calculate the total NLO correction in the 
Constrained Differential Renormalization (CDR) scheme. Analysis of the results, along 
with the increasing experimental precision,  
shows that it is feasible that the corrections at the Next-to-Next-to-Leading Order (NNLO) level 
may be important for the next generation of high-precision experiments.
\end{abstract}


\maketitle

\section{Introduction}

The M{\o}ller scattering measurements are not only one of the oldest and the best-established tools
of modern physics, but also a clean, powerful probe of New Physics (NP) effects~\cite{2}. 
More recently, the significant interest to this process from both theoretical and experimental communities has been re-ignited by
two precision experiments: E-158~\cite{e158} at SLAC, which made the first observation of parity violation in electron-electron scattering, 
and  MOLLER experiment planned at JLab \cite{JLab12}.
Both are dedicated to measuring the Parity-Violating (PV) 
asymmetry in the  $e^-e^- \rightarrow e^-e^- (\gamma)$ scattering at low energies. 
The MOLLER experiment aims to measure the PV asymmetry in the scattering of longitudinally
polarized electrons off unpolarized electrons with a combined statistical and systematic
uncertainty of 36 parts per billion~\cite{JLab12}. With the estimated systematic
contribution of 1\%, the measurement will still be statistics-limited,
and further improvements in precision might be possible with
additional running time. 
The measurement of the electron's weak charge
$Q_W^e$ with a 2.3\% accuracy planned by MOLLER would yield the most precise single
measurement of the weak mixing angle $\sin^2\theta_W$, with a fractional
accuracy of 0.1\%, at an average momentum transfer
$Q^2=0.0056~\mathrm{GeV}^2$. At this precision, MOLLER can shed light on the discrepancy between the
had\-ro\-nic and leptonic determinations of $\sin^2\theta_W$ at the $Z$-boson pole. 
Furthermore, the difference between the values of
$\sin^2\theta_W$ determined at the $Z$-boson pole and at low $Q^2$ is
sensitive to the NP effects at TeV scales. 

Before physics of interest can be extracted
from the experimental data, radiative effects must be carefully treated. 
MOLLER's stated precision goal is significantly
more ambitious than that of its predecessor E-158, 
so very precise theoretical input for this measurement will be crucial. 
In spite of significant  earlier theoretical effort dedicated to calculations 
of Electroweak Radiative Corrections (EWC) 
(see the early review papers \cite{Mo_Tsai_69} and \cite{Maximon69}, 
more recent \cite{Erler2005} and \cite{Kaiser2010}, 
and numerous additional references in our paper \cite{abiz1}), 
we believe that a new level of accuracy is required for the next-generation, 
high-precision experiments.
To match the expected experimental systematic uncertainty, it is desirable to keep
the theoretical uncertainty due to the radiative corrections at or
below the $0.1\%$ level. Obviously, calculating large sets of one-loop Feynman diagrams by hand is a tedious task. 
Recently, program packages such as FeynArts \cite{int3}, FormCalc \cite{Hahn}, LoopTools 
\cite{Hahn} and FORM \cite{int7} have created the possibility of handling the substantial number of 
diagrams reasonably quickly, minimizing probability of human errors, and preventing the rapid 
error accumulation often unavoidable with purely numerical methods. 
One of the key features of the presented work is to compare, step by step, the complete one-loop set 
of EWC to the PV Moller scattering asymmetry calculated first by hand and then with FeynArts, 
FormCalc and LoopTools as base languages using two different renormalization conditions.

FeynArts is a Mathematica package which provides the generation and visualization of Feynman 
diagrams and amplitudes involving Standard Model particles. FormCalc, a Mathematica package, 
reads diagrams generated with FeynArts and evaluates amplitudes with the help of the program 
FORM in analytical form. FORM, a successor to SCHOONSCHIP \cite{int8}, is a Symbolic Manipulation System which is also 
essential for our computer algebra-based method, and is used by FormCalc as a core program.
LoopTools provides the many-point tensor coefficient functions and is used to numerically 
evaluate scalar and tensor one-loop integrals.  
In FormCalc, it is possible for the regularization 
to be done either by dimensional reduction or by the usual dimensional regularization 
scheme. 
After that, one may implement one of the two renormalization schemes (RS), the on-shell scheme or 
the Constrained Differential Renormalization (CDR)scheme.
For calculations done at the one-loop level, the CDR scheme is equivalent to regularization done by dimensional reduction in  
the $\rm \overline{MS}$ scheme with redefined scale $\log {\bar M}^2 = \log \mu^2 +2$, 
where  $\bar M$ and $\mu$ are the renormalization scales in the CDR and in the dimensional regularization method, respectively \cite{Hahn}.

A complete automatization would limit the range of applications, so these packages are not 
"black box"; they require considerable human input on many stages. We call our approach based 
on FeynArts, FormCalc, LoopTools, and FORM "semi-automated". On the other hand, these packages 
allow modifications to better suit specific projects. In \cite{CM}, for example, 
we adopted FeynArts and FormCalc for the NLO calculations of the differential cross section 
in electron-nucleon scattering. 
In general, the results obtained with these packages can be presented in both analytical and 
numerical form. However, our equations for the EWC to the scattering asymmetry obtained with FeynArts 
and FormCalc are too lengthy and cumbersome, so we present only approximate equations 
obtained by hand. However, as we show in the numerical analysis section, the agreement between 
numerical results obtained with the two methods -- "by hand" and semi-automatic -- is excellent.

An additional way to ensure that our NLO EWC calculations are perfectly correct is the detailed 
comparison of results calculated with different renormalization conditions within the same scheme.  
Of course, the sum of all radiative corrections forming a full gauge-invariant set must be independent  
on the choice of renormalization conditions.
Paper \cite{DS1992}, for example, clearly demonstrated the cancellation 
of gauge dependencies in one-loop corrections from self-energies, tadpoles, vertex and box 
diagrams to physical amplitudes for four-fermion processes.
However, the agreement between the results evaluated in different renormalization schemes can be 
guaranteed only if we take into account all orders of perturbation expansion, not just NLO. 
Since in this article we are only dealing with one-loop corrections, we do not expect the results produced 
in different schemes to be identical. In fact, the difference we see between the results obtained 
with the on-shell and CDR schemes indicates the need to consider higher-order corrections. 
A detailed discussion was given in \cite{HT}, for example.
The NNLO corrections will be our next task.

For now, we concentrate on achieving 
the best accuracy possible in one selected scheme and perform NLO calculations using two methods 
and two sets of renormalization conditions.
For that, we choose the on-shell renormalization scheme with two different 
sets of renormalization conditions, 
the approach proposed by W.~Hollik in   
\cite{Hol90} (see also \cite{BSH86}) and the approach suggested by
A.~Denner in \cite{Denner}. For brevity, we will call Hollik's
renormalization conditions HRC and Denner's conditions DRC. 
It is obvious that any renormalization scheme is required to meet
physical conditions, although it is
possible to vary renormalization conditions for the sake of simplicity
of the problem and still keep the final gauge invariant results unchanged. 
As a result, contributions to the cross section coming from the
different non-gauge invariant loop corrections (self-energies and
vertex corrections) could vary greatly depending on the choice of
renormalization conditions.

The first goal of this paper is to calculate the full set of
one-loop EWC, both numerically with no simplifications using
semi-automatic approach, and "by hand", analytically in a compact asymptotic form 
\cite{7,8,9}, and compare the results. 
Our second goal is to present comparison and analysis of
the various contributions to the cross-section asymmetry calculated
within the HRC and DRC renormalization conditions.
Our third aim is to calculate the total NLO corrections in  
the CDR scheme and estimate the importance of the NNLO corrections for such high-precision experiments as MOLLER.

The rest of the paper is organized as follows. 
In Section~\ref{sec2} we provide details of the basic notation, the lowest-order (Born or Leading Order (LO))
and NLO contributions to M{\o}ller scattering. 
The same section gives a short description of photon emission which is essential for  
 removal of nonphysical parameters from regularized infrared divergent cross section. 
The details of the HRC and DRC renormalization conditions and a discussion of gauge 
invariance can be found in Section~\ref{sec3}. 
Analysis of analytical and numerical results in the on-shell RS using HRC and DRC renormalization conditions 
is given in the beginning of Section~\ref{sec4}. 
Later, in the same section, the CDR results are discussed. 
Section~\ref{sec5} includes the analysis of possible effects of 
an additional new-physics massive neutral boson on the observable asymmetry. 
Our conclusions and future plans are discussed in Section~\ref{sec:conclusion}.

\section{DEFINITIONS AND FRAMEWORK}
\label{sec2}

\begin{figure}
\resizebox{0.45\textwidth}{!}{%
  \includegraphics{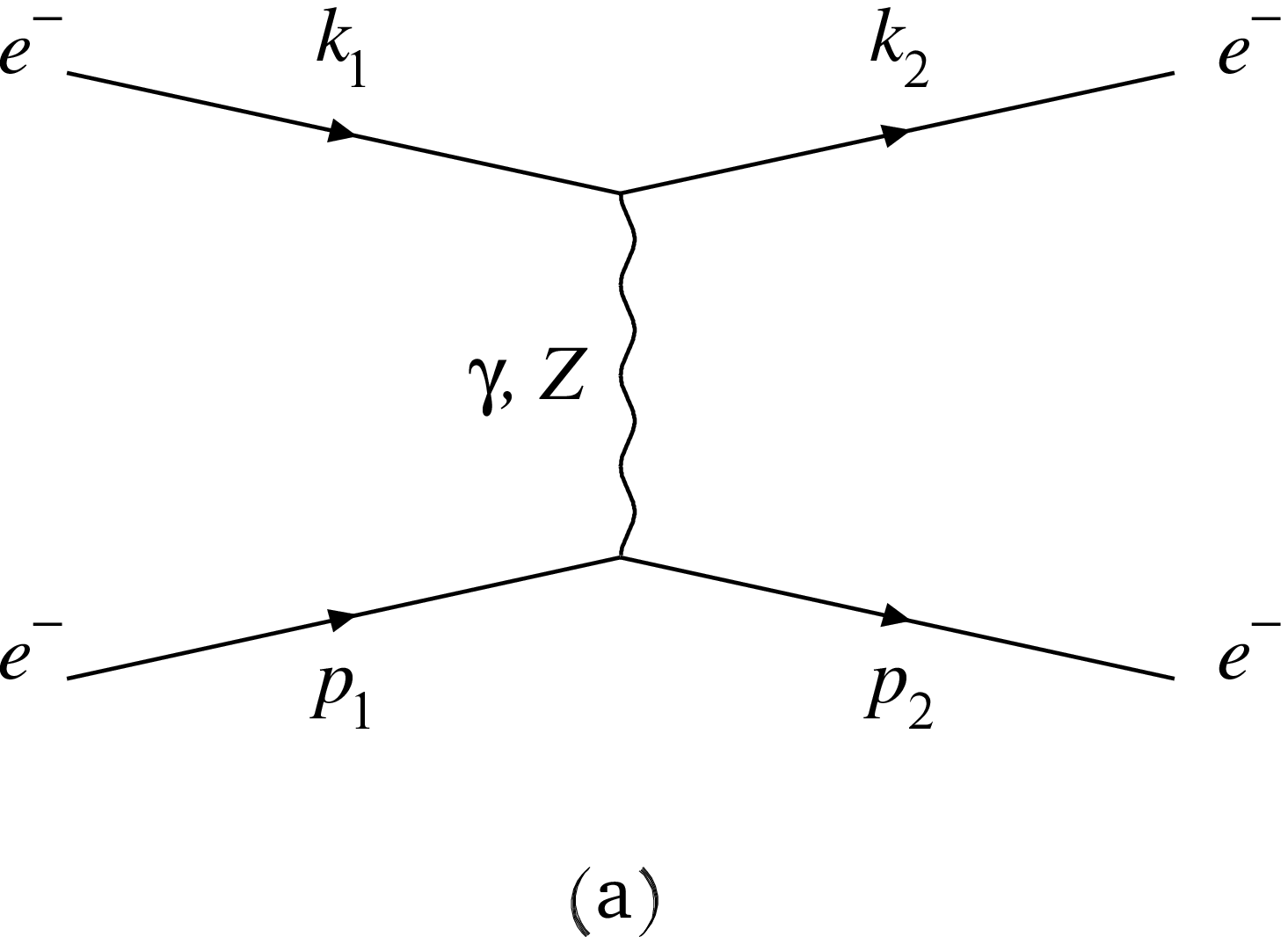}
\hspace*{2cm}       
  \includegraphics{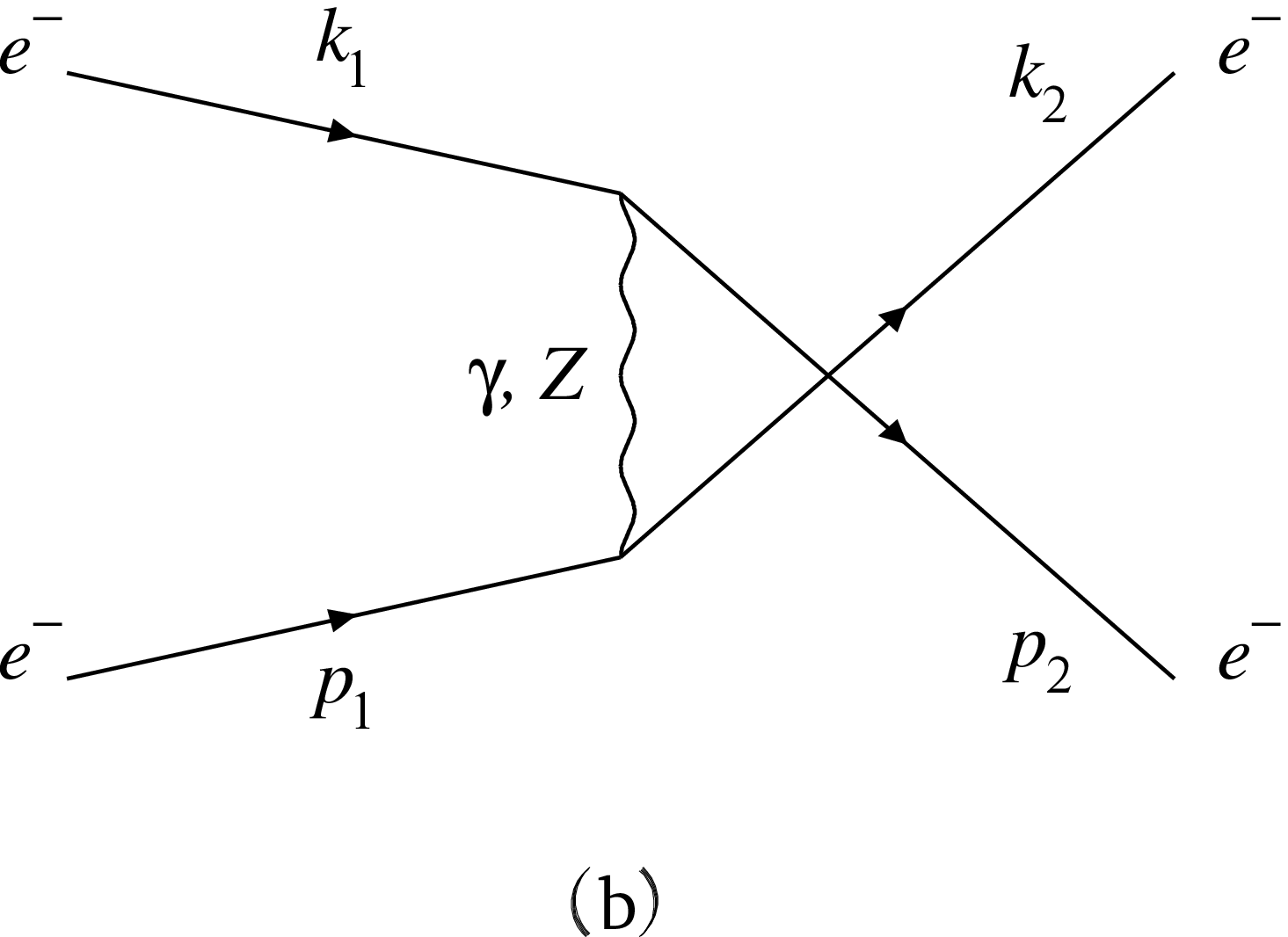}
}
\vspace*{0cm}       
\caption{Diagrams describing nonradiative M{\o}ller scattering in the (a) t- and (b) u-channels.}
\label{born}       
\end{figure}

In the Standard Model, the Born cross section for
 M{\o}ller scattering with the longitudinally-polarized electrons
\begin{eqnarray}
e^-(k_1)+e^-(p_1) \rightarrow e^-(k_2)+e^-(p_2)
\label{0} 
\end{eqnarray}
can be represented in the form
\begin{eqnarray}
\sigma^0 =\frac{\pi \alpha^2}{s}
\sum_{i,j=\gamma,Z} [\lambda_-^{i,j}(u^2D^{it}D^{jt}+t^2D^{iu}D^{ju})
\nonumber \\ 
   + \lambda_+^{i,j}s^2(D^{it}+D^{iu})(D^{jt}+D^{ju})],
\label{cs0} 
\end{eqnarray}
where
$\sigma \equiv {d\sigma}/{d \cos \theta}$ and 
$\theta$  is the scattering angle of the detected electron
with momentum $k_2$ in the center of mass system of the initial electrons. 
The set of momenta of initial ($k_1$ and $p_1$) and final
($k_2$ and $p_2$) electrons (see Fig.~\ref{born}) generates the standard
set of Mandelstam variables,
\begin{eqnarray}
s=(k_1+p_1)^2,\ t=(k_1-k_2)^2,\ u=(k_2-p_1)^2.
\label{stu}
\end{eqnarray}
We neglect the electron mass $m$ whenever possible and in particular when $m^2 \ll s,-t,-u$.

A useful structure we employ in this paper is
\begin{eqnarray}
D^{ir}=\frac{1}{r-m_i^2}\ \ (i=\gamma,Z;\ r=t,u),
\label{structure}
\end{eqnarray}
which depends on the $Z$-boson mass $m_Z$ or 
on the photon mass $m_\gamma \equiv \lambda $. The photon mass is set to zero
everywhere with the exception of specially indicated
cases where the photon mass is taken to be an
infinitesimal parameter that regularizes an infrared
divergence. 
In addition, we use the functions 
\begin{eqnarray}
{\lambda_{\pm}}^{i,k} =
     {\lambda_1}_B^{i,k}{\lambda_1}_T^{i,k} \pm {\lambda_2}_B^{i,k}{\lambda_2}_T^{i,k},
\label{b10}
\end{eqnarray}
which are combinations of coupling constants and the degrees of polarizations $p_{B(T)}$ 
of the electrons with momentum $k_1$ ($p_1$) given by
\begin{eqnarray}
 {\lambda_1}_{B(T)}^{i,j} = \lambda_V^{i,j} -p_{B(T)} \lambda_A^{i,j},\
   {\lambda_2}_{B(T)}^{i,j} = \lambda_A^{i,j} -p_{B(T)} \lambda_V^{i,j},
\nonumber
\end{eqnarray}
\begin{eqnarray}
 \lambda_V^{i,j}=v^iv^j + a^ia^j,\
   \lambda_A^{i,j}=v^ia^j + a^iv^j.
\end{eqnarray}
Here, vector and axial-vector parts of the couplings have the following structure
\begin{eqnarray}
 v^{\gamma}=1,\ a^{\gamma}=0,\
\nonumber \\ 
 v^Z=(I_e^3+2s_{W}^2)/(2s_{W}
c_{W}),
\ a^Z=I_e^3/(2s_{W}c_{W}).
\end{eqnarray}
It should be recalled that $ I_e^3=-1/2 $ and 
$s_{W}\  (c_{W})$ 
are the sine (cosine) of the Weinberg mixing angle which is defined in terms of $m_{Z}$ and $m_{W}$ 
according to the rules of the Standard Model (SM):
 $c_{W}=m_{Z}^2/m_{W}^2$ and $s_{W}=\sqrt{1-c_{W}^2}$.
The electron degrees of polarization $p_{B(T)}$ are labeled such that the subscripts
$L$ and $R$  correspond to the values of $p_{B(T)}$ = $-1$ and $p_{B(T)}$ = $+1$ respectively. Here, the first subscript
indicates the degree of polarization for the 
momentum $k_1$, while the second indicates the
degree of polarization for the momentum $p_1$. 
Combining the degrees of electron beam polarizations,
we can obtain four measurable cross sections. However by the virtue of the rotational invariance, 
two of them are identical: $\sigma_{LR}=\sigma_{RL}$. The three polarization cross sections
can be used to construct three independent asymmetries \cite{CC96}. Of particular interest to us is 
the parity-violating  asymmetry $A_{LR}$ which is defined as follows
\begin{eqnarray}
A_{LR} =
 \frac{\sigma_{LL}+\sigma_{LR}-\sigma_{RL}-\sigma_{RR}}
      {\sigma_{LL}+\sigma_{LR}+\sigma_{RL}+\sigma_{RR}}
 =
 \frac{\sigma_{LL}-\sigma_{RR}}
      {\sigma_{LL}+2\sigma_{LR}+\sigma_{RR}}.
\label{A}
\end{eqnarray}
This single-polarization asymmetry corresponding
to the scattering of longitudinally polarized electrons  on unpolarized electrons
is proportional to the
combination $1-4s_{W}^2$, and is therefore highly sensitive to small changes in $s_{W}$.
That is why the asymmetry $A_{LR}$ was used as the observable in E-158
and will be measured in the future MOLLER  experiment.
At low energies and at Born level, the PV asymmetry $A_{LR}^0$ is given by
\begin{eqnarray}
A_{LR}^0 = \frac{s}{2m_{W}^2}
\frac{y(1-y)}{1+y^4+(1-y)^4} \frac{1-4s_{W}^2}{s_{W}^2},\ \ y=-t/s.
\end{eqnarray}

\begin{figure*}
\resizebox{0.99\textwidth}{!}{%
  \includegraphics{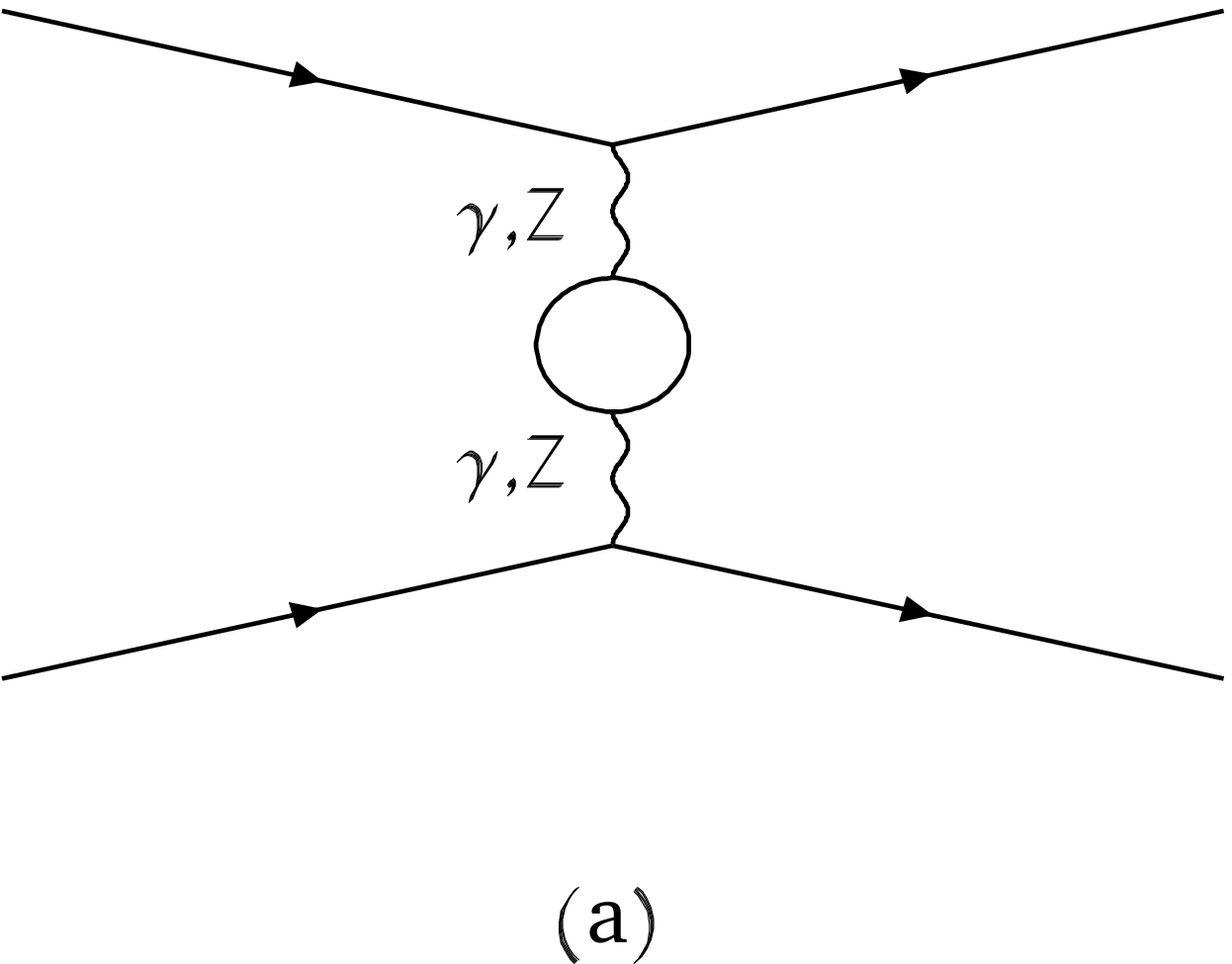}
\hspace*{0.95cm}       
  \includegraphics{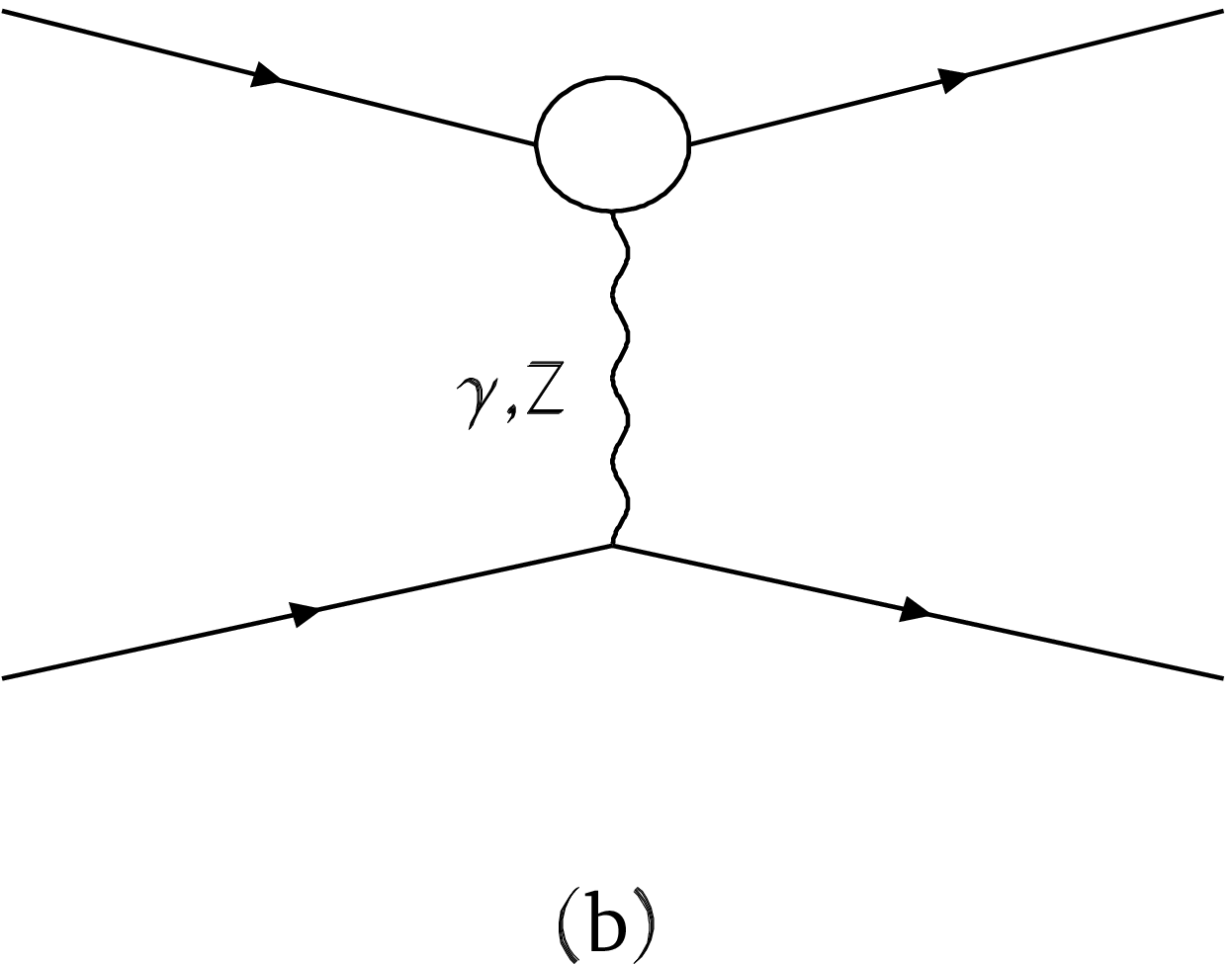}
\hspace*{0.95cm}       
  \includegraphics{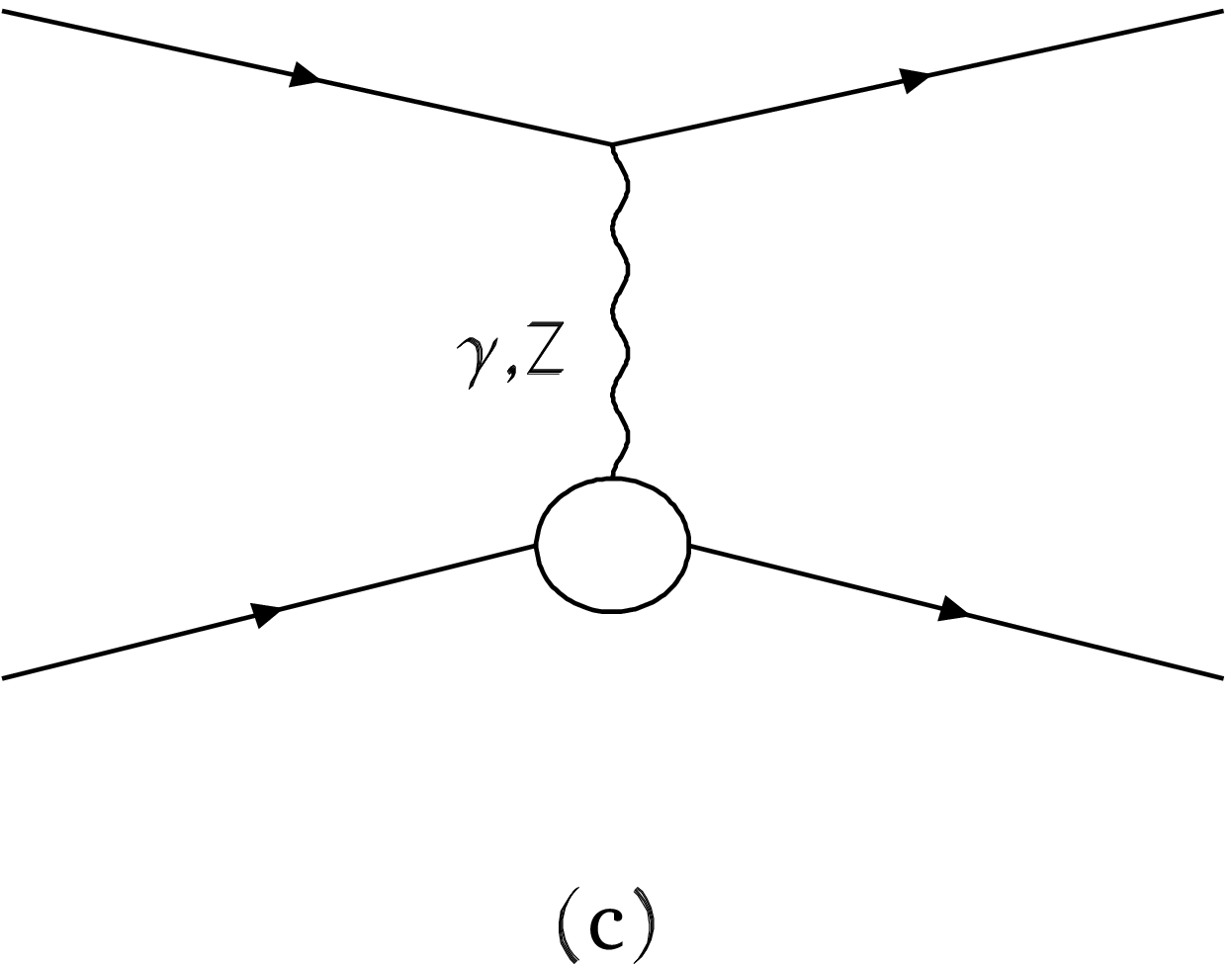}
\hspace*{0.95cm}       
  \includegraphics{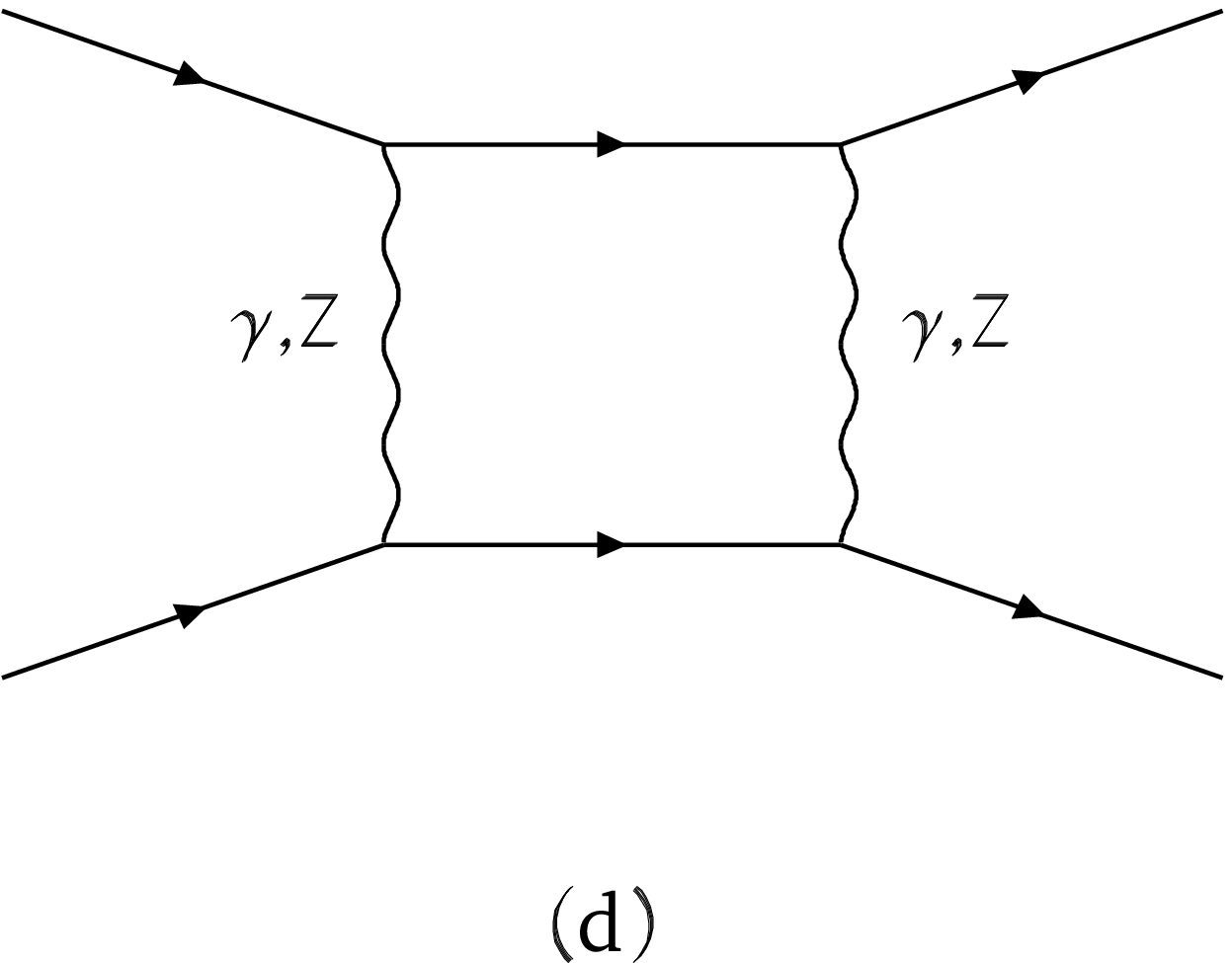}
\hspace*{0.95cm}       
  \includegraphics{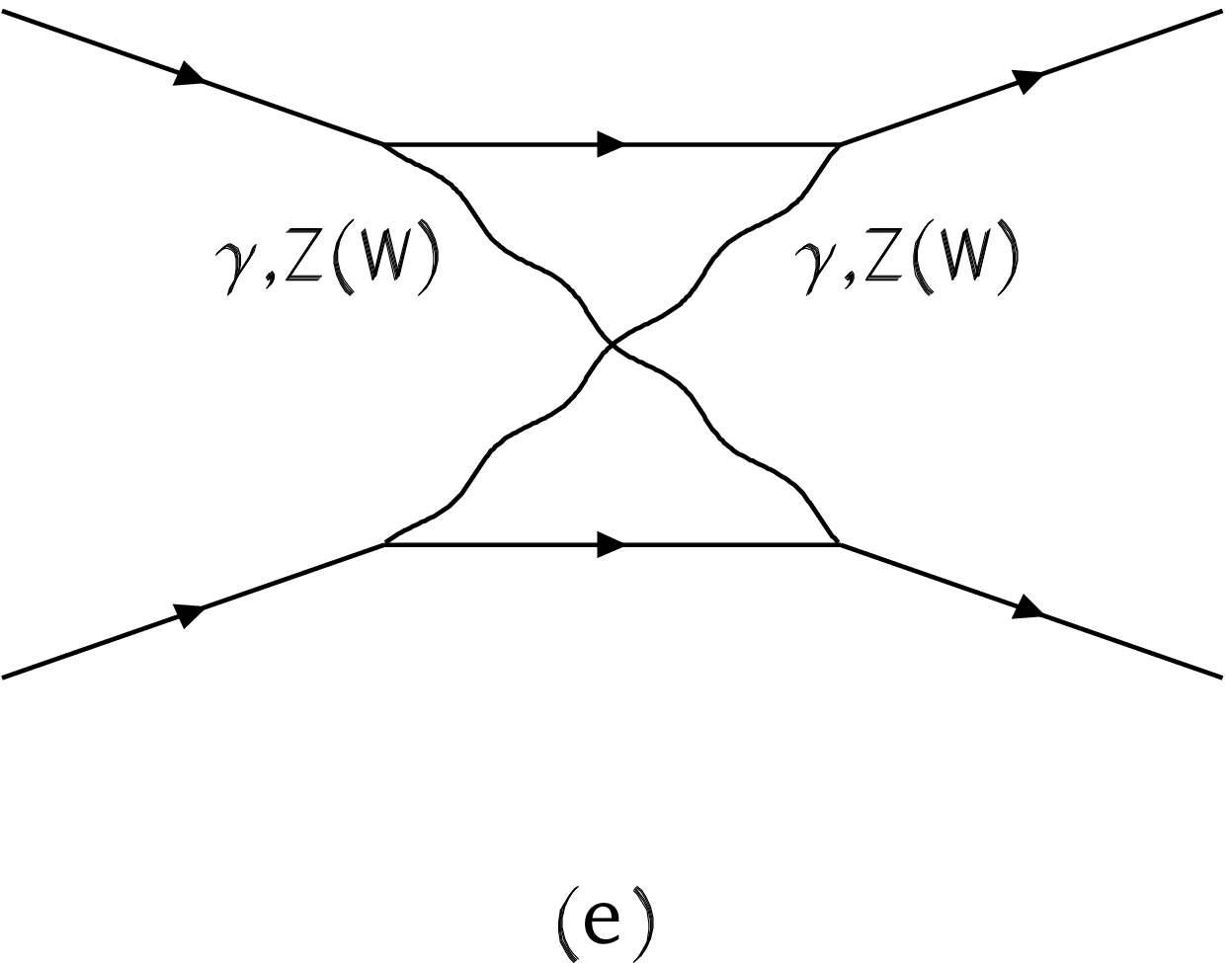}
}
\vspace*{1cm}       
\caption{One-loop t-channel diagrams for the M{\o}ller process.
The circles represent the contributions of self-energies
and vertex functions. 
The u-channel diagrams are obtained via the interchange $k_2 \leftrightarrow p_2$. }
\label{2f}      
\end{figure*}

The contribution of virtual particles ($V$-contribution) to the cross section 
of M{\o}ller scattering is described by the three classes of diagrams: 
boson self-energies (BSE) 
(they include $\gamma\gamma$, $\gamma Z$ and $ZZ$ self-energies and
are shown symbolically in Fig.~\ref{2f}(a)), 
vertex functions (Fig.~\ref{2f}(b) and \ref{2f}(c)), 
and two-boson exchange diagrams (boxes) shown in Fig.~\ref{2f}(d, e). 
In the on-shell and CDR renormalization schemes there is no contribution from the electron self-energies. 
The corresponding cross section is given by the sum
\begin{eqnarray}
\sigma^{V}=
\sigma^{\rm BSE}+\sigma^{\rm {Ver}} +\sigma^{\rm Box}.
\label{G}
\end{eqnarray}
The detailed expressions for all the terms in this sum were given in our recent paper \cite{abiz1}.

Contributions coming from the vertex correction graphs (with a photon in the loop), as well as the
$\gamma\gamma$ and $\gamma Z$ boxes suffer from the well-known infrared divergence. 
Regularization of this divergence can be done by giving the photon a small unphysical mass $\lambda$.
Obviously, the final result should be free of unphysical parameters and hence such  dependence has to be removed. 
That can be done if we consider additional contributions associated with photon emission diagrams (bremsstrahlung).
The detailed description of this contribution is also given in \cite{abiz1}.
The bremsstrahlung cross section can be broken
down into two parts (soft and hard) as 
\begin{eqnarray}
\sigma^{R}=
\sigma_{\rm {IR}}^{R}+\sigma_{H}^{R}
\label{broke}
\end{eqnarray}
by separating the integration domain according to $k_0 < \omega$ and $k_0 > \omega$, where
$k_0$ is the photon energy (in the reference frame co-moving
with the center of mass of the primary electrons). The parameter 
$\omega$ corresponds to the maximum of the emitted soft-photon energy. 
First, we follow the methods of paper \cite{HooftVeltman} to get a well-known result (see also \cite{5-DePo,7,8,9,6-Pe})
for the soft-photon cross section (where $e$ is the base of the natural logarithm):
\begin{eqnarray}
\sigma_{\rm {IR}}^{R}=
\frac{\alpha}{\pi} (4\log \frac{2\omega}{\lambda} \log \frac{tu}{em^2s}
-\log^2\frac{s}{em^2}
\nonumber \\ 
+1-\frac{\pi^2}{3} +\log^2\frac{u}{t}   )
\sigma^0.
\label{IRR}
\end{eqnarray}

Next, we sum the IR-terms of $V$- and $R$-contributions,
\begin{eqnarray}
\sigma^C=
\sigma_{\rm {IR}}^{V}+\sigma_{\rm {IR}}^R=
\frac{\alpha}{\pi} (4\log \frac{2 \omega}{\sqrt{s}} \log \frac{tu}{em^2s}
\nonumber \\ 
-\log^2\frac{s}{em^2}
+1-\frac{\pi^2}{3} +\log^2\frac{u}{t}   )
\sigma^0.
\label{can}
\end{eqnarray}
and get a result which is free of regularization parameter $\lambda$.  

At this point let us continue with the discussion of the details of renormalization conditions  
we use in our calculations.

\section{RENORMALIZATION CONDITIONS AND GAUGE INVARIANCE}
\label{sec3}

To obtain the ultraviolet-finite result and render the parameters of the Standard
Model real, 
 we have to apply a renormalization procedure. 
For a gauge-invariant set, physical results should be invariant under different 
 renormalization conditions. That is, although the contributions of the different types of diagrams can 
vary strongly for different renormalization conditions, the total impact of all one-loop virtual effects on observable 
quantities must remain independent.   
 In other words, the contributions of separate self-energies and vertex correction functions 
 strongly depend on the details of the renormalization conditions, and to properly account 
of the EWC they should be taken as one gauge-independent set. We will illustrate this for the case 
of the observable $A_{LR}$, which is especially sensitive to the renormalization conditions.
In addition, we can verify that our results are correct by comparing the computer-based (DRC) and 
"by hand" (HRC) calculations.
We now briefly describe our two chosen renormalization conditions, DRC and HRC,  within the on-shell renormalization scheme. 

Both use multiplicative renormalization constants, and as a result the electroweak Lagrangian, 
originally written in terms of bare parameters, is separated into a basic Lagrangian and 
a counterterm Lagrangian. 
The basic Lagrangian has the same form as the bare one, but depends
on renormalized parameters and fields. The counterterm Lagrangian depends
on renormalization constants of masses, charges and fields. Renormalization
constants are fixed by the renormalization conditions, which are separated
into two classes: the first determines the renormalization of the parameters,
and the second fixes the renormalization of fields. The first class is related
to  physical observables at a given order of perturbation theory, and
the second one is related to the Green's functions and has no effect
on calculations of $S$-matrix elements. Both approaches use essentially
the same renormalization conditions to fix the parameters of the SM in the following way:
\begin{eqnarray}
& \mbox{Re}\hat{\Sigma}_{T}^{W} (m_{W}^{2} )=
  \mbox{Re}\hat{\Sigma}_{T}^{Z} (m_{Z}^{2} )=
  \mbox{Re}\hat{\Sigma}^{f}     (m_{f}^{2} )=0,
\nonumber\\
& \hat{\Gamma}_{\mu}^{ee\gamma}\left(k^{2}=0,\, p^{2}=m^{2}\right)=ie\gamma_{\mu}.
\label{eq:dh1}
\end{eqnarray}

\begin{figure*}
\resizebox{0.99\textwidth}{!}{%
  \includegraphics{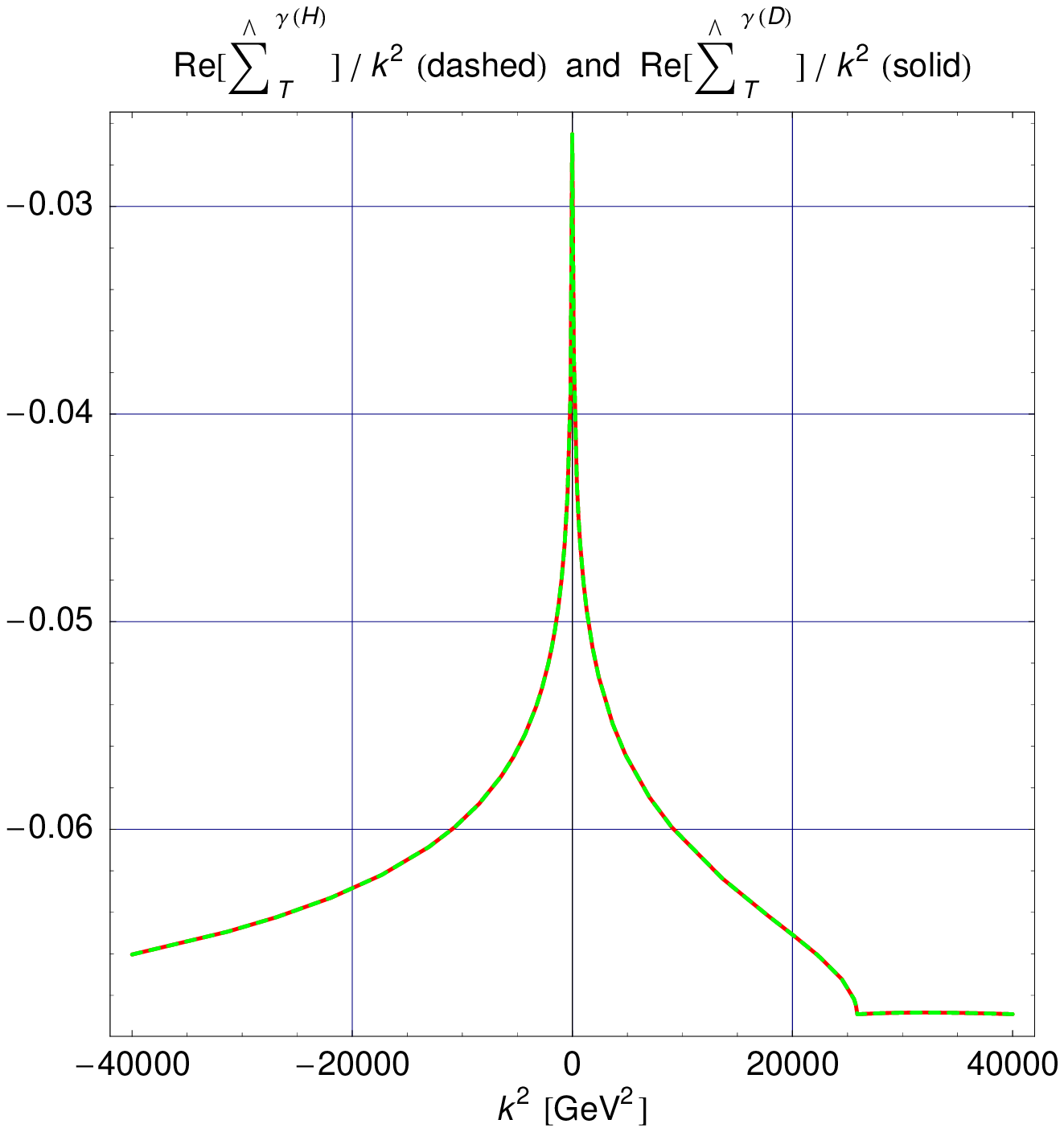}
\hspace*{0.95cm}       
  \includegraphics{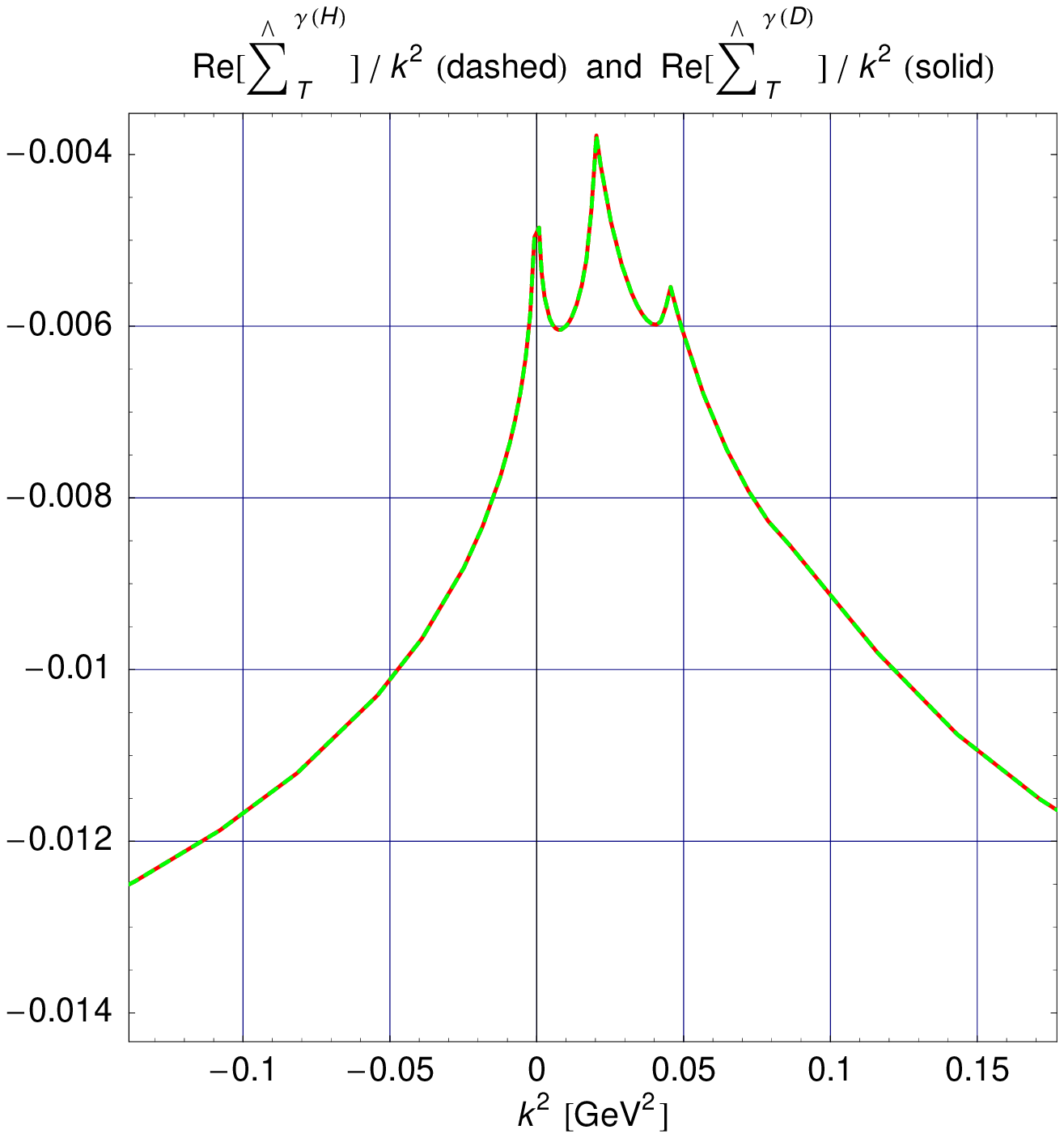}
}
\vspace*{1cm}       
\caption{Truncated and renormalized $\gamma\gamma$ self-energies in both
sets of renormalization conditions. The right graph shows the low-energy domain.
}
\label{gr1}      
\end{figure*}

\begin{figure*}
\resizebox{0.99\textwidth}{!}{%
  \includegraphics{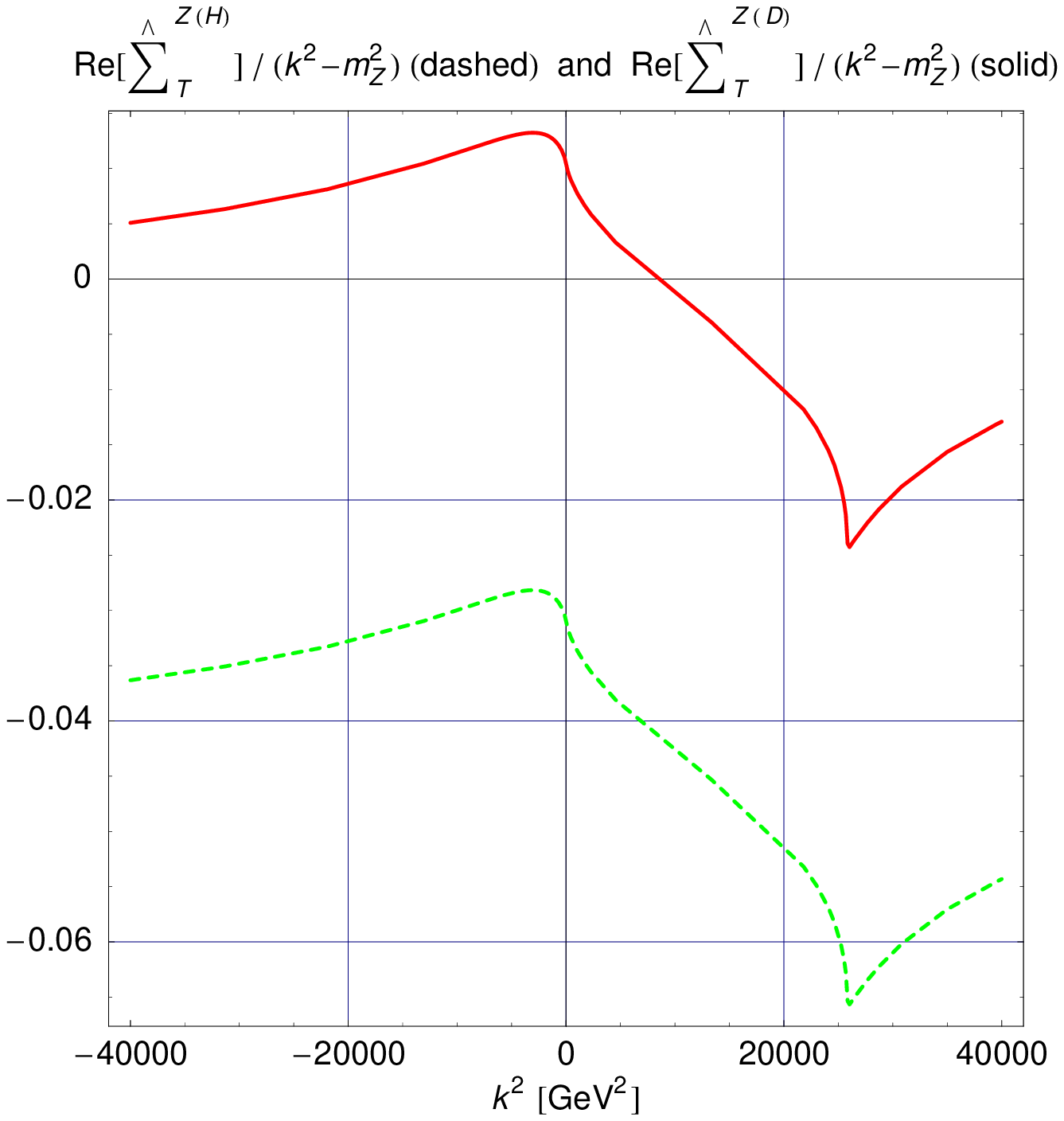}
\hspace*{0.95cm}       
  \includegraphics{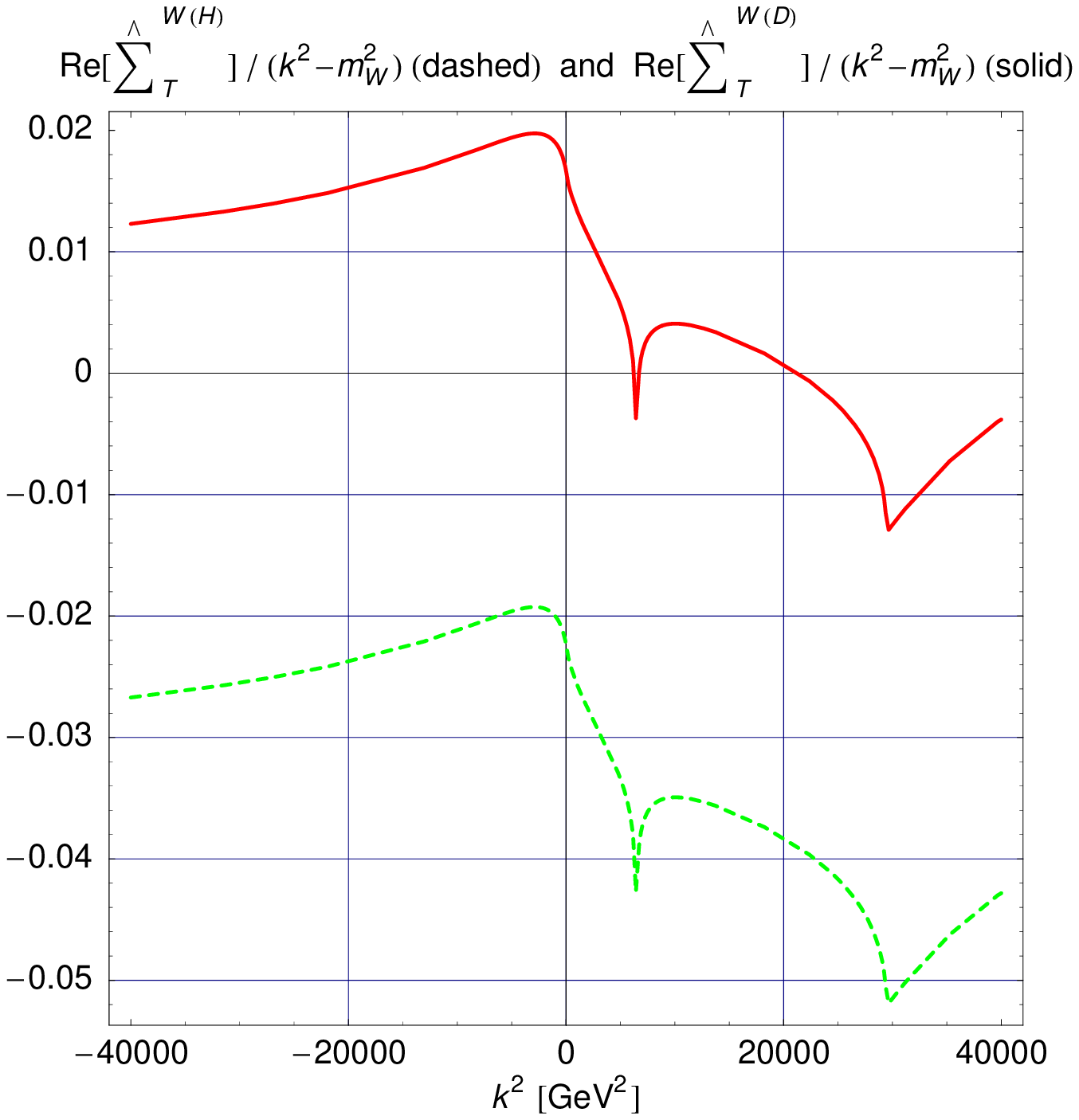} 
}
\vspace*{1cm}       
\caption{Truncated and renormalized $ZZ$ and $WW$ self-energies
in both sets of renormalization conditions.
}
\label{gr2}      
\end{figure*}

\begin{figure*}
\resizebox{0.99\textwidth}{!}{%
  \includegraphics{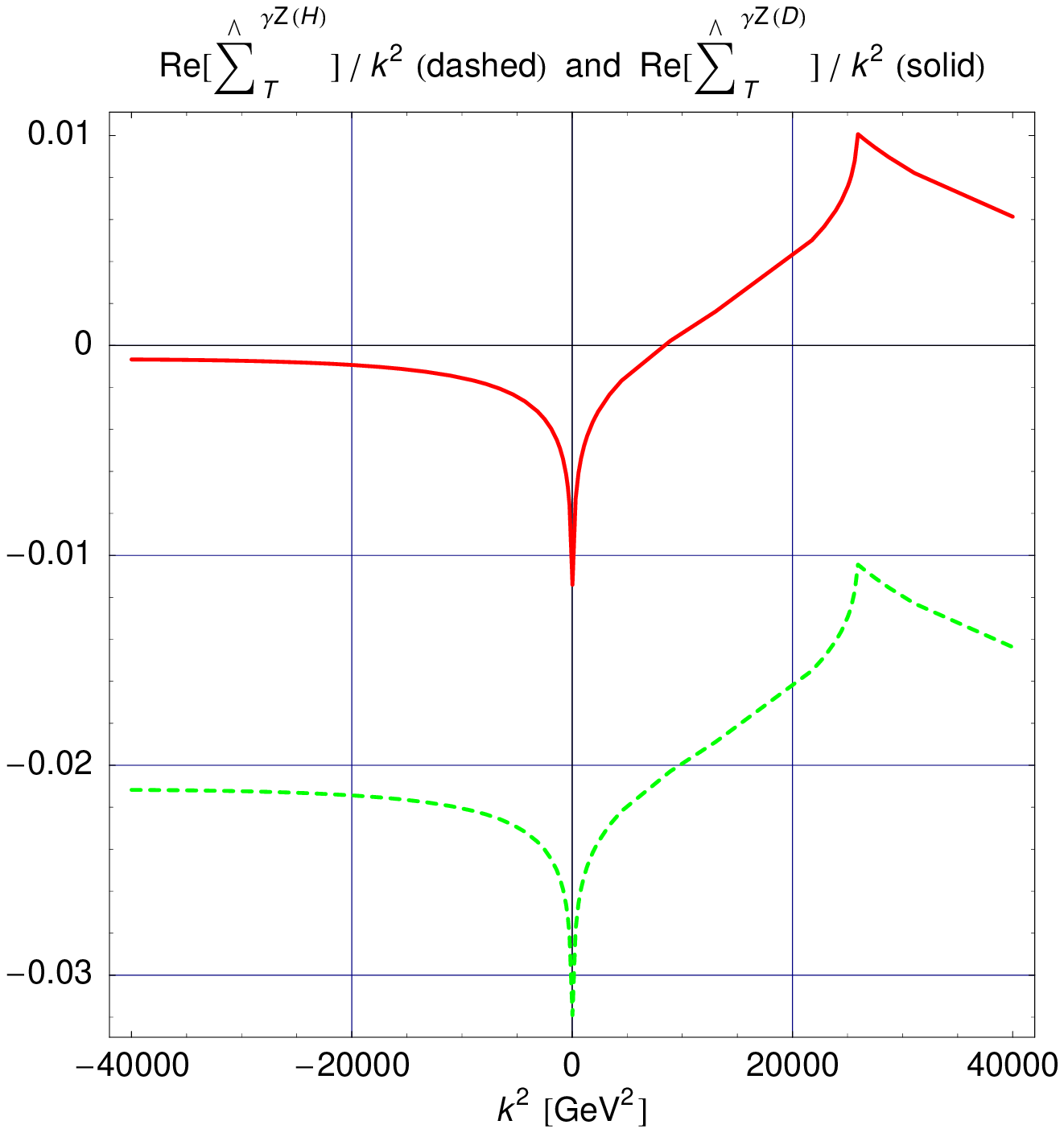}
\hspace*{0.95cm}       
  \includegraphics{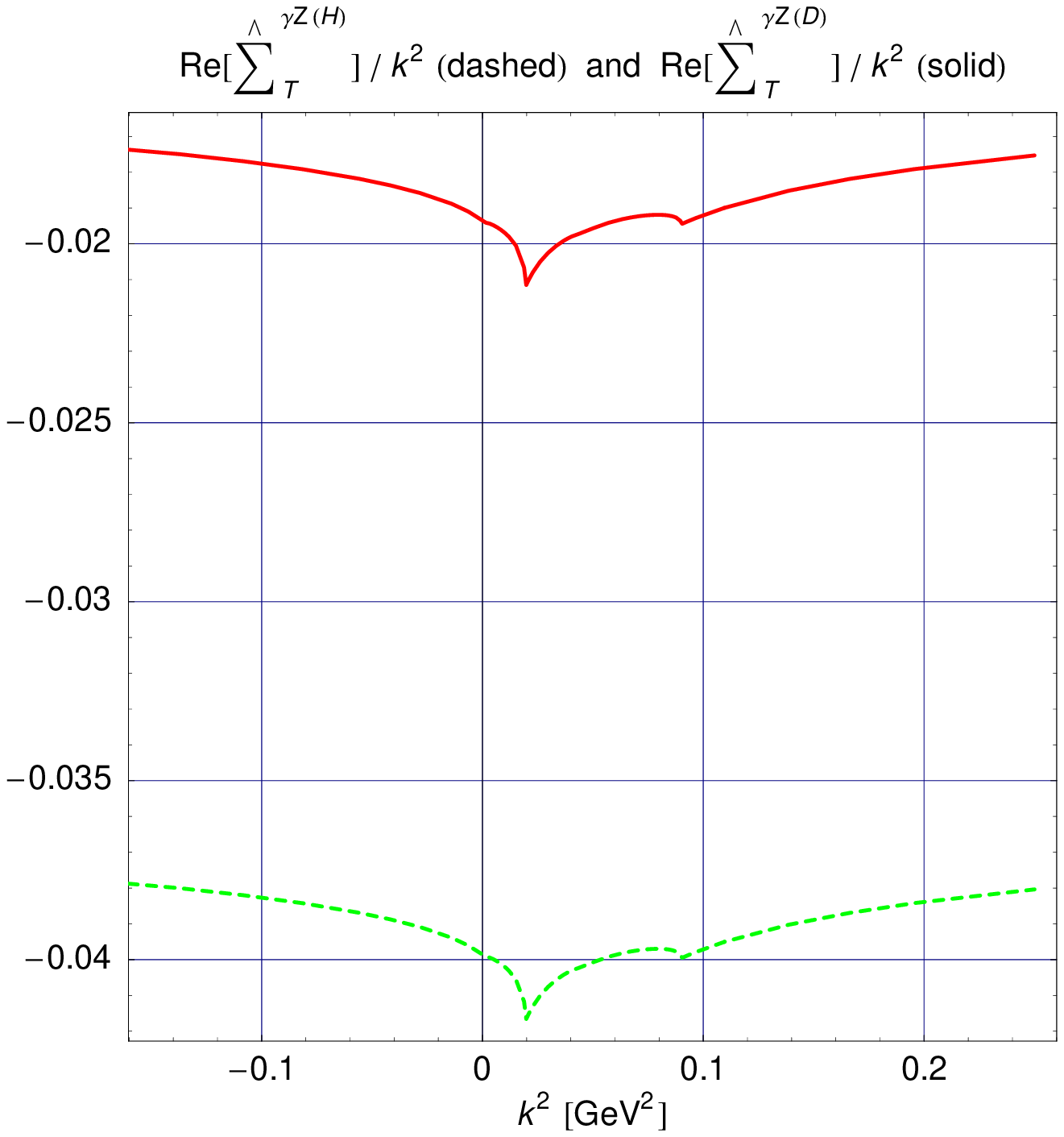}
}
\vspace*{1cm}       
\caption{Truncated and renormalized $\gamma Z$ self-energies
in both sets of renormalization conditions.
The right graph shows the low-energy domain.
}
\label{gr2B}      
\end{figure*}

Here, $\mbox{Re}\hat{\Sigma}_{T}^{Z, W}\left(m_{Z, W}^{2}\right)$ 
and $ \hat{\Gamma}_{\mu}^{ee\gamma}\left(k^{2}=0,\, p^{2}=m^{2}\right)$ 
are the real parts of the truncated, transverse renormalized boson self-energy and electron 
vertex correction graphs, respectively. The longitudinal parts of the boson self-energy make very small contributions
 and are not considered here.
The first condition of Eq.~(\ref{eq:dh1}) fixes the mass renormalization of the $W$-, $Z$-bosons and fermions
without quark mixing. The second condition fixes the renormalization of electric charge, 
and is derived from the Thomson limit when momentum transfer $k^{2}=0$ and external electrons are on their
mass shell. As for the renormalization conditions of the fields, both approaches are quite different.
In HRC, field renormalization constants are
determined from the following conditions:
\begin{eqnarray}
\displaystyle
 & \hat{\Sigma}_{T}^{\gamma Z}\left(0\right)=0,\  \ \
\frac{\partial}{\partial k^{2}}\hat{\Sigma}_{T}^{\gamma}\left(0\right)=0.
\label{eq:dh2}
\end{eqnarray}
However, in the DRC renormalization conditions, the field renormalization
is defined on-shell, as it was done for renormalization of
the SM parameters. This explicitly introduces an additional set of conditions,
besides Eq.~(\ref{eq:dh1}) and Eq.~(\ref{eq:dh2}), which
read:
\begin{eqnarray}
\mbox{Re}\hat{\Sigma}_{T}^{\gamma Z}\left(m_{Z}^{2}\right)=0,\ 
\mbox{Re}\frac{\partial}{\partial k^{2}}\hat{\Sigma}_{T}^{Z}\left(m_{Z}^{2}\right)=0,\ 
\nonumber \\ 
 \mbox{Re}\frac{\partial}{\partial k^{2}}\hat{\Sigma}_{T}^{W}\left(m_{W}^{2}\right)=0.
\label{eq:dh3}
\end{eqnarray}
As a result, in DRC, renormalization constants
for the fields of vector bosons are calculated in a relatively simple way, 
without the mass-renormalization constants: 
\begin{eqnarray}
&&\delta Z_{W}^{(D)}=-\mbox{Re}\frac{\partial}{\partial k^{2}}\Sigma_{T}^{W}\left(m_{W}^{2}\right),\
\nonumber \\
&&\delta Z_{Z}^{(D)}=-\mbox{Re}\frac{\partial}{\partial k^{2}}\Sigma_{T}^{Z}\left(m_{Z}^{2}\right),
\nonumber \\
&&\delta Z_{Z\gamma}^{(D)}=\frac{2}{m_{Z}^{2}}\mbox{Re}\Sigma_{T}^{\gamma Z}\left(0\right),\ 
 \delta Z_{\gamma Z}^{(D)}=-\frac{2}{m_{Z}^{2}}\mbox{Re}\Sigma_{T}^{\gamma Z}\left(m_{Z}^{2}\right),\
\nonumber\\
&& \delta Z_{\gamma}^{(D)}=-\frac{\partial}{\partial k^{2}}\Sigma_{T}^{\gamma}\left(0\right).
\label{eq:dh4} 
\end{eqnarray}
They can be presented through truncated and non-re\-normalized self-energy graphs. 
In comparison with HRC, where the renormalization conditions
of Eq.~(\ref{eq:dh3}) are not present, field renormalization constants
are defined in a different way and depend on the mass-renormalization
constants:
\begin{eqnarray}
&& \delta Z_{\gamma}^{(H)}=-\frac{\partial}{\partial k^{2}}\Sigma_{T}^{\gamma}\left(0\right),
\nonumber\\
&& \delta Z_{Z}^{(H)}=\frac{\partial}{\partial k^{2}}
\Sigma_{T}^{\gamma}\left(0\right)-2\frac{c_{W}^{2}-s_{W}^{2}}{s_{W}c_{W}}
\frac{\Sigma_{T}^{\gamma Z}\left(0\right)}{m_{Z}^{2}}
\nonumber\\
&&+2\frac{c_{W}^{2}-s_{W}^{2}}{s_{W}^{2}}
\left(\frac{\delta m_{Z}^{2}}{m_{Z}^{2}}-\frac{\delta m_{W}^{2}}{m_{W}^{2}}\right),
\nonumber\\
&& \delta Z_{W}^{(H)}=\frac{\partial}{\partial k^{2}}\Sigma_{T}^{\gamma}
\left(0\right)-2\frac{c_{W}}{s_{W}}\frac{\Sigma_{T}^{\gamma Z}\left(0\right)}{m_{Z}^{2}}
\nonumber\\
&&+\frac{c_{W}^{2}}{s_{W}^{2}}\left(\frac{\delta m_{Z}^{2}}{m_{Z}^{2}}-\frac{\delta m_{W}^{2}}{m_{W}^{2}}\right),
\nonumber\\
&& \delta Z_{Z\gamma}^{(H)}=\frac{c_{W}s_{W}}{c_{W}^{2}-s_{W}^{2}}
\left(\delta Z_{Z}^{(H)}-\delta Z_{\gamma}^{(H)}\right).
\label{eq:dh5}
\end{eqnarray}
The presence of the mass renormalization constants in the field-renormalization
Eq.~(\ref{eq:dh5}) increases the values
of the truncated and renormalized self-energy diagrams, and
the dominant NLO contributions to the observable cross
section come from these diagrams. In DRC,
the mass renormalization constants appear in renormalization
constants of the electroweak couplings, and hence we observe
comparable contributions coming from both self-energies and vertex
corrections. Of course, such a comparison has no physical meaning since
neither self-energies nor vertex corrections represent a gauge-invariant
set on their own. As is well known, only the sum of both groups is gauge invariant;
later we show that both
approaches give exactly the same results for the observable asymmetry.
 We would like to highlight that it is important
to exercise caution when comparing separate contributions arising
from the different renormalization conditions. This point is illustrated
by Fig.~\ref{gr1}, ~\ref{gr2} and ~\ref{gr2B},  where one can see various renormalized vector
boson self-energies calculated with both DRC and HRC.

Let us note that the same renormalization conditions are imposed on the electromagnetic
field. As we can see from Fig.~\ref{gr1}, which shows the truncated and renormalized $\gamma\gamma$ self-energies, 
there is no difference whatsoever between the two sets of conditions. The situation
is quite different if we look at the results for the truncated and renormalized $ZZ$, $\gamma Z$ and $WW$ self-energies. 
In Fig.~\ref{gr2} and ~\ref{gr2B}, we can see a substantial difference in the results obtained 
within the two sets of renormalization conditions, 
where the DRC set systematically leads to the self-energies being smaller in magnitude.
As a result, with DRC, the self-energy contributions to M{\o}ller asymmetry are roughly a factor of two smaller in value compared to the values given by
HRC. However, adding the DRC vertex corrections restores the total  correction obtained with DRC to within 0.001\% of the HRC result
at all energies relevant to the planned MOLLER experiment at JLab.

Now we are ready to proceed to the analysis of the results.
Section~\ref{sec41} is done in a manner independent of the renormalization conditions, and all 
notations that follow below can be applied to both DRC and HRC of the on-shell scheme.

\section{RESULTS AND ANALYSIS}
\label{sec4}

\subsection{On-Shell Renormalization}
\label{sec41}

For the numerical analysis, we use 
$\alpha=1/137.035 999$,\ 
$m_W=80.399\ \mbox{GeV}$,\  and
$m_Z=91.1876\ \mbox{GeV}$ according to \cite{PDG10}, take electron, muon, and $\tau$-lepton 
masses as $m_e=0.510 998 910\ \mbox{MeV}$,  $m_\mu=0.105 658 367\ \mbox{GeV}$, $m_\tau=1.776 84\ \mbox{GeV}$, 
and quark masses for loop contributions as
$m_u=0.069 83\ \mbox{GeV}$,\ $m_c=1.2\ \mbox{GeV}$,\ $m_t=174\ \mbox{GeV}$,
$m_d=0.069 84\ \mbox{GeV}$,\ $m_s=0.15\ \mbox{GeV}$, and $m_b=4.6\ \mbox{GeV}$.
The light quark masses provide $\Delta \alpha_{had}^{(5)}(m_Z^2)$=0.02757 \cite{jeger}, 
where
\begin{eqnarray}
\Delta \alpha_{had}^{(5)}(s)=\frac{\alpha}{\pi} \sum_{f=u,d,s,c,b} Q_f^2 \biggl(\log\frac{s}{m_f^2}-\frac{5}{3}\biggr),
\end{eqnarray}
$Q_f$ is the electric charge of fermion $f$ in proton charge units $q,\ (q=\sqrt{4\pi \alpha})$.
We can see that the use of the light quark masses 
as parameters regulated by the had\-ro\-nic vacuum polarization
is a better choice in this case. 
We checked that variations of the light quark masses around the outlined values have a negligible 
impact on the values of the polarization asymmetry, so the choice of quark masses does not introduce 
a significant uncertainty to our results. 
An earlier work by \cite{Erler2005}, relevant to observables measured at very low momentum transfers, 
which determined the weak mixing angle in the $\rm \overline{MS}$-scheme, 
also argued that the uncertainty from non-perturbative had\-ro\-nic contributions is small compared to 
anticipated experimental uncertainties. Although \cite{6-Pe} argued 
that the most significant source of theoretical uncertainty on scattering asymmetry comes from the 
had\-ro\-nic contributions to the $\gamma Z$ vacuum polarization, we find that 
in our calculations had\-ro\-nic contributions are under good control.    
Finally, for the mass of the Higgs boson, we take $m_H=115\ \mbox{GeV}$. 
Although this mass is still to be determined experimentally, the dependence of the EWC on $m_H$ is rather weak.

Let us determine the physical impact of this contribution to the observable $A_{LR}$,
by defining the relative corrections 
to the Born asymmetry as
$$\delta^{\rm C}_A = (A_{LR}^{\rm C}-A_{LR}^0)/A_{LR}^0, $$
where the index C stands for a specific contribution, for example $\rm C=BSE, Ver, Box$.
Let indicies $\gamma$SE, $\gamma Z$SE and $Z$SE denote $\gamma\gamma$-, $\gamma Z$-, and $ZZ$-BSE contributions,
respectively. 
The main subject of the following analysis is "{\it weak}" relative corrections, which are defined as all BSE contributions
(including the $\gamma\gamma$-SE which is not weak by nature, but is needed here 
to account for all IR-finite contributions to the asymmetry) plus
heavy vertex (HV) contributions ("heavy" means "massive", i.e. $Z$- or $W$-boson), 
$ZZ$- and $WW$-boxes. In summary: {\it weak} = BSE+HV+$ZZ$+$WW$.

\subsubsection{ Analysis of BSE contributions to PV asymmetry}

We start with the $Z$SE-contribution, where
\begin{eqnarray}
\delta_A^{Z\rm SE} &&= \frac{A_{LR}^{Z\rm SE}-A_{LR}^0}{A_{LR}^0}
=\frac{ \frac{(\sigma^0+\sigma^{Z\rm SE})|_{\rm LL-RR}}{\sigma^0_{00}+\sigma^{Z\rm SE}_{00}} 
            - \frac{\sigma^0|_{\rm LL-RR}}{\sigma^0_{00}} }
      { \frac{\sigma^0|_{\rm LL-RR}}{\sigma^0_{00}} }
\nonumber \\
&&\approx
\frac{ \sigma^{Z\rm SE}|_{\rm LL-RR} }{ \sigma^0|_{\rm LL-RR} }.
\label{addit}
\end{eqnarray}                                 
The operation $E|_{\rm LL-RR}$ under expression $E$  means $E_{LL}-E_{RR}$. 
Subscript $00$ denotes the unpolarized cross section.
The approximate equality is possible because $\sigma^{Z\rm SE}_{00}/\sigma^0_{00}$ is very small.
The denominator of the last fraction is calculated directly from Eq. (\ref{cs0}):
\begin{eqnarray}
\sigma^0|_{\rm LL-RR} &&= 16 \pi \alpha^2 v^Za^Z s (D^{\gamma t}+D^{\gamma u})(D^{Z t}+D^{Z u})
\nonumber \\
&&\approx - 32 \pi \alpha^2 v^Za^Z  \frac{s}{m_Z^2} (D^{\gamma t}+D^{\gamma u}).
\end{eqnarray}                     
With simplifications, the numerator of Eq.~(\ref{addit}) is
\begin{eqnarray}
\sigma^{Z\rm SE}|_{\rm LL-RR} 
 && \approx \frac{\pi \alpha^2}{s} D_S^{ZZ t}  
( D^{\gamma t} M_{\rm ev}^{Z\gamma Z \gamma}
\nonumber \\
&&- D^{\gamma u}  M_{\rm od}^{Z\gamma Z \gamma} )|_{\rm LL-RR} + (t \leftrightarrow u) 
\nonumber \\
&& 
\approx
- 16 \pi \alpha^2 v^Za^Z \frac{s}{m_Z^4} (D^{\gamma t}+D^{\gamma u})
\nonumber \\
&& 
\times ( \hat \Sigma_T^Z(t)+\hat \Sigma_T^Z(u) ).
\end{eqnarray}                                 
Finally,
\begin{eqnarray}
\delta_A^{Z\rm SE} \approx \frac{ \hat \Sigma_T^{Z}(t)+\hat \Sigma_T^{Z}(u) }{2m_Z^2}.
\end{eqnarray}                                 
At small $r$ (Eq. (\ref{structure})
) corresponding to $E_\mathrm{lab}=11$~GeV and $\theta=90^{\circ}$, the corrections are
$\delta_A^{Z\rm SE}$ (HRC) $\approx 0.0309$ vs.
$\delta_A^{Z\rm SE}$ (DRC) $\approx -0.0105$.

Similarly, for $\gamma Z$SE-contribution
\begin{eqnarray}
\sigma^{\gamma Z\rm SE}|_{\rm LL-RR} 
\approx 
\frac{\pi \alpha^2}{s} D_S^{\gamma Z t} [ 
  D^{\gamma t} (M_{\rm ev}^{\gamma\gamma Z \gamma} + M_{\rm ev}^{Z\gamma\gamma \gamma})
\nonumber \\
- D^{\gamma u} (M_{\rm od}^{\gamma\gamma Z \gamma} + M_{\rm od}^{Z\gamma\gamma \gamma})
]|_{\rm LL-RR} + (t \leftrightarrow u) \approx
\nonumber \\
 \approx
 16 \pi \alpha^2 a^Z \frac{s}{m_Z^2} 
 (D^{\gamma t}+D^{\gamma u})
\Bigl( \frac{\hat \Sigma_T^{\gamma Z}(t)}{t}+\frac{\hat \Sigma_T^{\gamma Z}(u)}{u} \Bigr).
\end{eqnarray}                                 
and
\begin{eqnarray}
\delta_A^{\gamma Z\rm SE} \approx -\frac{1}{2v^Z}
\Bigl( \frac{\hat \Sigma_T^{\gamma Z}(t)}{t}+\frac{\hat \Sigma_T^{\gamma Z}(u)}{u} \Bigr).
\end{eqnarray}                                 
For $\gamma Z$SE
at $E_\mathrm{lab}=11$~GeV and $\theta=90^{\circ}$, the corrections are 
$\delta_A^{\gamma Z\rm SE}$ (HRC) $\approx -0.6028$ vs.
$\delta_A^{\gamma Z\rm SE}$ (DRC) $\approx -0.2909$.
It is important to note that 
the deviation in $\hat \Sigma_T^{\gamma Z}$ has a dramatic impact on such a sensitive 
observable as $A_{LR}$. For example, the uncertainty in $\hat \Sigma_T^{\gamma Z}$
of 1\% will result in a change in $\delta_A^{\gamma Z\rm SE}$ of up to 0.05.

All terms with properties of Eq.~(\ref{addit}) contribute additively to the total
correction, for example,
\begin{eqnarray}
\delta_A^{\gamma Z {\rm SE} + Z \rm SE} \approx 
\delta_A^{\gamma Z \rm SE } +
\delta_A^{ Z \rm SE}. 
\end{eqnarray}                                 
We call such contributions {\it additive}. 
$\gamma$SE gives a {\it non-additive} and small contribution that we consider later.

\subsubsection{  Analysis of HV and box contribution to PV asymmetry}

Starting with the $\Lambda_2$-contribution, which 
comes from the triangle diagrams with an additional massive boson, $Z$ or $W$,  
we get
\begin{eqnarray}
\delta_A^{\Lambda_2} \approx
\frac{ \sigma^{\Lambda_2}|_{\rm LL-RR} }{ \sigma^0|_{\rm LL-RR} }.
\end{eqnarray}                                 
The numerator, with some approximations, is
\begin{eqnarray}
\sigma^{\Lambda_2}|_{\rm LL-RR} 
\approx 
8 \alpha^3 v^Za^Z s (D^{\gamma t}+D^{\gamma u})
\nonumber \\
\times
\Bigl( \frac{\Lambda_2(t,m_Z)}{t} + \frac{\Lambda_2(u,m_Z)}{u} \Bigr),
\end{eqnarray}                                 
so the correction is proportional to $\Lambda_2$ in the following way:
\begin{eqnarray}
\delta_A^{\Lambda_2} \approx -\frac{\alpha m_Z^2}{4\pi}
\Bigl( \frac{\Lambda_2(t,m_Z)}{t} + \frac{\Lambda_2(u,m_Z)}{u} \Bigr).
\end{eqnarray} 
                                
In HRC, we can  simplify the result by
using series expansion of $\Lambda_2$ at small $t$:
\begin{eqnarray}
\Lambda_2(t,m_Z)=-\frac{t}{3m_Z^2} \bigl( 2\log\frac{-t}{m_Z^2}-\frac{23}{6} \bigr) +{\cal O}(\frac{t}{m_Z^2}),
\end{eqnarray}                                 
which gives
\begin{eqnarray}
\delta_A^{\Lambda_2} \approx \frac{\alpha}{6\pi}(\log\frac{tu}{m_Z^4}-\frac{23}{6}).
\end{eqnarray}                                 
The numerical value obtained from above at $E_\mathrm{lab}=11$~GeV and $\theta=90^{\circ}$ gives $\delta_A^{\Lambda_2} \approx -0.0125$, which is in 
agreement with the exact (semi-automatic) numerical calculations.

The $\Lambda_3$-contribution, 
which represents the triangle diagrams with a 3-boson vertex, $WW\gamma$ or $WWZ$,
is calculated in a similar way, 
so we present only the final result:
\begin{eqnarray}
\delta_A^{\Lambda_3} \approx -\frac{3 \alpha m_Z^2}{32 \pi s_W^2 v^Za^Z}
\Bigl( \frac{\Lambda_3(t,m_W)}{t} + \frac{\Lambda_3(u,m_W)}{u} \Bigr).
\end{eqnarray}        
After simplifications and series expansion of $\Lambda_3$ at small $t$,
\begin{eqnarray}
\Lambda_3(t,m_W)=-\frac{5 t}{27 m_W^2} + {\cal O}(\frac{t}{m_W^2}),
\end{eqnarray}                                 
we find
\begin{eqnarray}
\delta_A^{\Lambda_3} \approx \frac{\alpha}{\pi} \frac{5}{9(1-4s_W^2)}.
\label{eq43}
\end{eqnarray}                                 
Using Eq.~(\ref{eq43}) for $E_\mathrm{lab}=11$~GeV and $\theta=90^{\circ}$, we obtain
$\delta_A^{\Lambda_3} \approx 0.0118$. Again, this approximate value calculated "by hand" is in 
a very good agreement with the exact result obtaned with our second, computer-based approach.

The box part is UV-finite and does not require the renormalization procedure.
We divide the box contribution into QED ($\gamma\gamma$- and $\gamma Z$-boxes) 
and a heavy-box part (HB = $ZZ+WW$):
\begin{eqnarray}
\sigma^{\rm Box}
= \sigma^{\rm Box}_{\rm QED} + \sigma^{\rm Box}_{\rm HB}.
\label{box}
\end{eqnarray}
The types of boxes are shown in Fig.~\ref{2f}(d, e).
The IR-divergent QED-part of boxes (the first term in Eq.~(\ref{box})) is described in detail 
both analytically and numerically in \cite{abiz1}.
For the purely-weak part of the boxes (the second term), the equations are derived in the low-energy approximation.

The total {\it weak} correction to $A_{LR}$ includes the HB cross section:
\begin{eqnarray}
\sigma_{\rm HB}^{\rm Box} =-\frac{\alpha^3}{s}
 \sum_{k=\gamma,Z} (B_{ZZ}^k+B_{WW}^k)  
+ (t \leftrightarrow u),
\end{eqnarray}
where the  expressions for $B_{ij}^k$ take a form
\begin{eqnarray}
&&B_{ZZ}^k=D^{kt}
\lambda_-^{B k} \delta^1_{ZZ} +
(D^{kt}+D^{ku})\lambda_+^{B k} \delta^2_{ZZ},
\nonumber \\[0.3cm] \displaystyle
&&B_{WW}^k=D^{kt}
\lambda_-^{C k} \delta^1_{WW} +
(D^{kt}+D^{ku})\lambda_+^{C k} \delta^2_{WW}.
\end{eqnarray}
The combinations of the coupling constants are given in Eq.~(\ref{b10}).
Let us recall the coupling constants for the heavy boxes: 
\begin{eqnarray}
v^B={(v^Z)}^2+{(a^Z)}^2,\ a^B=2v^Za^Z, 
\nonumber \\
v^C=a^C=1/(4s_W^2).
\end{eqnarray}
At $s,|t|,|u| \ll m_Z^2$, the corrections $\delta_{(ij)}^{1,2}$ have a form:
\begin{eqnarray}
        \delta^1_{ZZ} = \frac{3u^2}{2m_Z^2},\
        \delta^2_{ZZ} = - \frac{3s^2}{2m_Z^2},\
\nonumber \\
        \delta^1_{WW} = \frac{2u^2}{m_W^2},\
        \delta^2_{WW} = \frac{s^2}{2m_W^2}.
\label{del-box}
\end{eqnarray}

At last, after simplification at small $t$,
for the relative corrections to PV asymmetry coming from heavy boxes we find: 
\begin{eqnarray}
\delta_A^{ZZ} \approx -\frac{3\alpha}{2\pi} v^B,\
\delta_A^{WW} \approx \frac{\alpha}{4\pi s_W^2(1-4s_W^2)}.
\end{eqnarray}                                 
The numerical values obtained from the equations above at
$E_\mathrm{lab}=11$~GeV and \ $\theta=90^{\circ}$ give 
$\delta_A^{ZZ} \approx -0.0013$ and  
$\delta_A^{WW} \approx 0.0238$,  which is once again in good
agreement with the exact results evaluated with help of the FeynArts, FormCalc, LoopTools and Form program packages.

\subsubsection{ Numerical analysis on EWC to PV asymmetry}

In the table below, we present the contributions to  relative {\it weak} corrections calculated
using two different approaches. In the first approach, we use approximate and compact expressions derived "by hand" 
with the application of HRC. In the second, we use computer-based analytical (FeynArts, FormCalc, and FORM) and when 
numerical (LoopTools) calculations, with DRC.

\begin{table*}
\caption{
The Born asymmetry $A_{LR}^0$ and the structure of relative {\it weak} 
corrections to it for $E_\mathrm{lab}=11$ GeV at different $\theta$.
}
\label{tab:1}       
{\vspace*{5mm}
\begin{tabular}{|c||c|c|c|c|c|c|c|c|}
\hline
  \multicolumn{1}{|c||}{{$\theta$,$^{\circ}$}} 
& \multicolumn{1}{ c|}{20} 
& \multicolumn{1}{ c|}{30} 
& \multicolumn{1}{ c|}{40} 
& \multicolumn{1}{ c|}{50} 
& \multicolumn{1}{ c|}{60} 
& \multicolumn{1}{ c|}{70} 
& \multicolumn{1}{ c|}{80} 
& \multicolumn{1}{ c|}{90}  \\
\hline                                                                                                               
{$A_{LR}^0$, ppb}       &$    6.63 $&$   15.19 $&$   27.45 $&$   43.05 $&$   60.69 $&$   77.68 $&$   90.28 $&$   94.97 $\\ 
\hline                                                                                                               
{$\gamma\gamma$-SE}, DRC  &$ -0.0043 $&$ -0.0049 $&$ -0.0054 $&$ -0.0058 $&$ -0.0062 $&$ -0.0064 $&$ -0.0066 $&$ -0.0067 $\\
{$\gamma\gamma$-SE}, HRC  &$ -0.0043 $&$ -0.0049 $&$ -0.0054 $&$ -0.0058 $&$ -0.0062 $&$ -0.0064 $&$ -0.0066 $&$ -0.0067 $\\
\hline                                                                                                               
{$\gamma Z$-SE}, DRC      &$ -0.2919 $&$ -0.2916 $&$ -0.2914 $&$ -0.2912 $&$ -0.2911 $&$ -0.2910 $&$ -0.2909 $&$ -0.2909 $\\
{$\gamma Z$-SE}, HRC      &$ -0.6051 $&$ -0.6043 $&$ -0.6042 $&$ -0.6038 $&$ -0.6034 $&$ -0.6031 $&$ -0.6028 $&$ -0.6028 $\\
\hline                                                                                                               
{$ZZ$-SE}, DRC            &$ -0.0105 $&$ -0.0105 $&$ -0.0105 $&$ -0.0105 $&$ -0.0105 $&$ -0.0105 $&$ -0.0105 $&$ -0.0105 $\\
{$ZZ$-SE}, HRC            &$  0.0309 $&$  0.0309 $&$  0.0309 $&$  0.0309 $&$  0.0309 $&$  0.0309 $&$  0.0309 $&$  0.0309 $\\
\hline                                                                                                               
{HV}, DRC                 &$ -0.2946 $&$ -0.2633 $&$ -0.2727 $&$ -0.2703 $&$ -0.2714 $&$ -0.2712 $&$ -0.2711 $&$ -0.2710 $\\
{HV}, HRC                 &$ -0.0015 $&$ -0.0012 $&$ -0.0010 $&$ -0.0009 $&$ -0.0008 $&$ -0.0007 $&$ -0.0007 $&$ -0.0007 $\\
\hline                                                                                                               
{$ZZ$-box}, exact         &$ -0.0013 $&$ -0.0013 $&$ -0.0013 $&$ -0.0013 $&$ -0.0013 $&$ -0.0013 $&$ -0.0013 $&$ -0.0013 $\\
{$ZZ$-box}, approx.       &$ -0.0013 $&$ -0.0013 $&$ -0.0013 $&$ -0.0013 $&$ -0.0013 $&$ -0.0013 $&$ -0.0013 $&$ -0.0013 $\\
\hline                                                                                                               
{$WW$-box}, exact         &$  0.0239 $&$  0.0238 $&$  0.0238 $&$  0.0239 $&$  0.0239 $&$  0.0238 $&$  0.0238 $&$  0.0238 $\\
{$WW$-box}, approx.       &$  0.0238 $&$  0.0238 $&$  0.0238 $&$  0.0238 $&$  0.0238 $&$  0.0238 $&$  0.0238 $&$  0.0238 $\\
\hline                                                                                                               
total {\it weak}, DRC, exact   &$ -0.5643 $&$ -0.5430 $&$ -0.5508 $&$ -0.5489 $&$ -0.5500 $&$ -0.5495 $&$ -0.5493 $&$ -0.5493 $\\
total {\it weak}, HRC, approx. &$ -0.5526 $&$ -0.5514 $&$ -0.5511 $&$ -0.5505 $&$ -0.5500 $&$ -0.5496 $&$ -0.5493 $&$ -0.5493 $\\
\hline
\end{tabular}}
\end{table*}

Table 1 demonstrates that
the $\gamma\gamma$-SE contribution is small, {\it non-additive} and, as expected, is the same whether obtained in HRC or DRC.
The $\gamma Z$-SE, $ZZ$-SE and HV contributions are rather sizeable, are all additive, and are different for HRC and DRC.
The $ZZ$-box contribution is small, and the $WW$-box is dominant for the weak box correction. 
Both the $ZZ$-box and $WW$-box are additive and their sum is in excellent 
agreement regardless the method of calculations.
The total relative {\it weak} correction is significant and in excellent 
agreement between the different methods. That confirms that we are dealing with a gauge-invariant 
set of graphs.
The discrepancy between the two approaches is $\sim 0.0001$ at $\theta=90^{\circ}$, but becomes 
larger with decreasing $\theta$.

\begin{figure}
\resizebox{0.45\textwidth}{!}{%
  \includegraphics{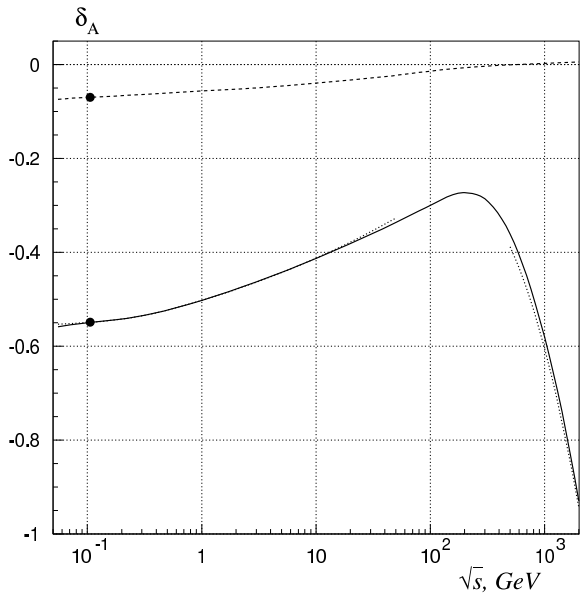}
}
\vspace*{0cm}       
\caption{The relative weak (solid line in DRC (semi-automated) and dotted line 
in HRC ("by hand")) and  QED (dashed line) corrections 
to the Born asymmetry $A_{LR}^0$ versus $\sqrt{s}$ at $\theta$ = 90$^{\circ}$. 
The filled circle corresponds to our predictions for the MOLLER experiment.}
\label{sverka2}       
\end{figure}

In Fig.~\ref{sverka2} we can see the relative {\it weak} corrections shown by solid line for DRC (exact) 
and dotted line for HRC (approximate). The dashed line shows the QED correction obtained by
including soft bremsstrahlung to the Born asymmetry $A_{LR}^0$. We can see that for low energy region $1 < \sqrt{s} < 30$~GeV the 
results calculated by the two methods are in excellent agreement. 
It is worth mentioning here that the semi-automated numerical calculations of boxes 
in the region of $\sqrt{s} \ll 1$ GeV suffer from the numerical instability due to Landau singularities. 
As for our approximated calculations, we have used the small-energy approximation with 
the expansion parameters taken as ${r}/{m_{Z,W}^2}$ for energies $\sqrt{s}<30$~GeV.
In any case, for the 11 GeV relevant for the planned JLab experiment, the consistency of
our calculations in both approaches is obvious, with a difference of $\sim 0.01$\% or less.
The dotted line for $ \sqrt{s} > 500$~GeV on the Fig.~\ref{sverka2} is obtained using HRC 
with the help of equations from \cite{36}, which used the high-energy approximation. 
We can see good a agreement between our results for the high-energy region $\sqrt{s} > 500$~GeV which becomes better with energy increase.
For $\sqrt{s} \geq $ 50 GeV we have excellent
agreement with the result of \cite{5-DePo} 
if we use their SM parameters (see \cite {abiz1}). Furthermore, the relative QED correction 
(see Fig. 8 in  \cite{5-DePo} and dashed line in Fig.~\ref{sverka2} here) 
is also in good qualitative and numerical agreement.
In this case, we apply the same cut on the soft photon emission energy as in  \cite{5-DePo} 
($\omega/\sqrt{s}=0.05$).
At the low-energy point corresponding to the E-158 experiment, and using our set of input parameters 
($\alpha$, $m_{W}$ and $m_{Z}$)  
we find that  $\delta_A^{weak}$ $\approx -54\%$. 
If we translate our input parameters to the set $\alpha$, $G_{F}$ and $m_{Z}$
according to \cite{5-DePo}, we obtain good agreement with the result of \cite{4-CzMa}.

\subsection{Constrained Differential Renormalization}

The CDR (Constrained Differential Renormalization) sche\-me, which provides renormalized expressions for 
Feynman graphs preserving the Ward identities, was introduced at the one-loop level in \cite{cdr2}. 
\cite{cdr1} expands on \cite{cdr2} to introduce the techniques for one-loop calculations 
in any renormalizable theory in four dimensions. The procedure has been implemented
in  FormCalc and LoopTools, which allows us to evaluate NLO EWC in CDR. 
Since our "scheme of choice" at the moment is on-shell, which is more suitable 
for calculating EWC beyond one-loop, we do not provide the same detailed analysis 
and step-by-step comparison between the two methods for CDR as we do for on-shell. 
The reason we evaluate NLO EWC in CDR is to obtain some indication of the size 
of the higher-order effects (NNLO and beyond) to see if there is enough motivation 
to do these very involved calculations in the future.

\begin{figure}
\resizebox{0.45\textwidth}{!}{%
  \includegraphics{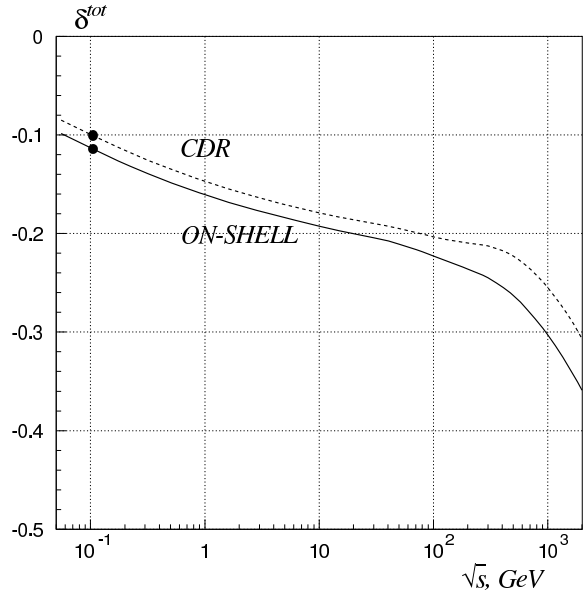}
}
\vspace*{0cm}       
\caption{The relative total corrections
to the unpolarized cross section versus $\sqrt{s}$ at $\theta$ = 90$^\circ$.
The filled circle corresponds to our predictions to the MOLLER experiment.  
Solid line corresponds to  CDR and dotted line to on-shell RS.
}
\label{sverka-xs-cdr}       
\end{figure}

In Fig.~\ref{sverka-xs-cdr}, we can see the relative total correction  
$$\delta^{\rm tot} = (\sigma^{\rm tot}-\sigma^0)/\sigma^0  $$
to the unpolarized cross section  versus $\sqrt{s}$ at $\theta$ = 90$^{\circ}$ 
for different RS: on-shell and  CDR.
In the region of small energies, the difference between the two schemes is almost constant 
and rather small ($\sim 0.01$), but grows at $\sqrt{s} \geq m_Z$.
It is well known that in the region of small energies, the correction to the cross section is dominated by the QED contribution. However, in the high-energy region the weak correction becomes comparable to QED. 
Since the difference between the on-shell and CDR results grows substantially as the weak correction becomes larger, it is  clear
that for an observable such as the PV asymmetry the difference between the on-shell and CDR schemes will be sizeable for the entire spectrum 
of energies $\sqrt{s} < 2000$~GeV.
Because of that, we expect that the NNLO correction to the PV asymmetry 
may become important to PV precision physics in the future.

\begin{figure}
\resizebox{0.45\textwidth}{!}{%
  \includegraphics{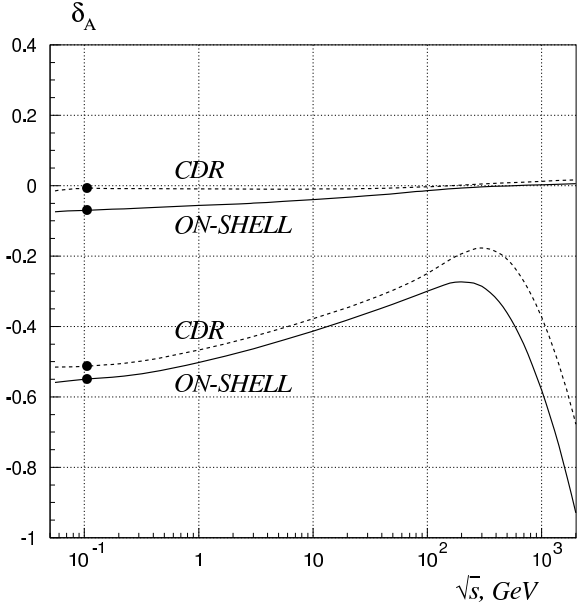}
}
\vspace*{0cm}       
\caption{The relative weak (lower lines) and  QED (upper lines) corrections to the Born 
asymmetry $A_{LR}^0$ versus $\sqrt{s}$ at $\theta$ = 90$^\circ$.
The filled circle corresponds to our predictions to the MOLLER experiment.  Solid lines 
correspond to  CDR and dotted lines to on-shell RS.
}
\label{sverka-as-cdr}       
\end{figure}

Fig.~\ref{sverka-as-cdr} shows the relative {\it weak} (lower lines), 
and  QED (upper lines) corrections  to the Born asymmetry $A_{LR}^0$ versus $\sqrt{s}$ at $\theta$ = 90$^\circ$. 
The difference is significant and is growing with increasing $\sqrt{s}$.
According to our calculations 
for $E_{lab}$ = 11 GeV, $\omega$ = 0.05 $\sqrt{s}$ and $\theta$ =
90$^\circ$, the total radiative correction to PV asymmetry is
$-69.8$\% with on-shell and $-58.5$\% with CDR.  
The difference is not at all surprising. For E-158, for example, the one-loop 
weak corrections were found to be about $-40$\% in the 
$\rm \overline{MS}$ scheme \cite{4-CzMa} and about $-50$\% in the on-shell scheme \cite{9,6-Pe}.

\begin{figure}
\resizebox{0.45\textwidth}{!}{%
  \includegraphics{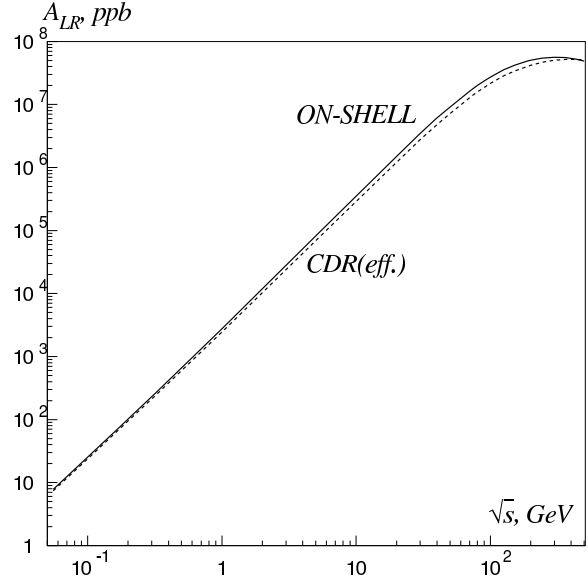}
}
\vspace*{0cm}       
\caption{The NLO-corrected asymmetries vs $\sqrt{s}$ at
  $\theta=90^\circ$, computed in on-shell RS (solid line) and CDR
  (dotted line). The CDR Born asymmetry uses the
  $\overline{\mathrm{MS}}$ definition of
  $\hat{s}_Z^2\equiv\sin^2\hat{\theta}_W(M_Z)=0.2313$~\cite{PDG10}.
}
\label{sverka-hat}       
\end{figure}

The physical, NLO-corrected asymmetries, computed in both on-shell and
CDR schemes, are compared in Fig.~\ref{sverka-hat}. Here, for
consistency with the $\overline{\mathrm{MS}}$ definition of the
couplings to $\mathcal{O}(\alpha)$~\cite{Sirlin1989}, we use
$\hat{s}_Z^2\equiv\sin^2\hat{\theta}_W(M_Z)=0.2313$~\cite{PDG10} 
in the expression of the Born asymmetry. We find that the predictions
for the physical PV asymmetry, computed to the same order in
perturbation theory in two different schemes, differ by about 3\%. The
difference is an indication of the order of magnitude the
higher-order, NNLO and beyond, terms. 

The \cite{6-Pe} estimated that the higher-order corrections are suppressed by $\sim0.1\%$ relative to the one-loop result, 
possibly 5\% in some cases, and thus are not significant source of uncertainty. 
However, we conclude that although the corrections at the NNLO level were not mandated 
by the previously achievable experimental precision, they may become important for 
the next generation of experiments.

\section{Effect of additional massive neutral boson}
\label{sec5}

Let us now add a very simple NP assumption to our SM calculations and show 
how this NP contribution affects the observable asymmetry. The reason we want to do it in here is to 
investigate if the two complimentary methods we used in the previous sections, "by-hand" and semi-automated, 
can be applied in the NP domain. As we mention in the Introduction, FeynArts, FormCalc, LoopTools, and FORM 
are not "black box" programs and can be modified for specific projects, including adding the NP sector. 
As was already concluded in  \cite{Mus2003} and \cite{Mus2008}, the proposed MOLLER 
measurement could be influenced by radiative loop effects of 
new-physics particles. This type of calculation is out of scope of this paper, but we plan to provide a full estimation in our future publication. 
For now, we assume that there is just one additional neutral boson (ANB), or  $Z'$-boson,  
with the usual $V-A$ structure of interaction with fermions, vector(axial)
coupling constants $v^{Z'}(a^{Z'})$ and  mass  $m_{Z'}$.
From the analysis done in the previous section, we can clearly see that in the 
low-energy region where $s,|t|,|u| \ll m_Z^2 < m_{Z'}^2$,  contributions 
are mainly suppressed by propagator factors like $D^{Z'r}$. 
In this section, our goal is to analyze the contribution of $Z'$-Born and $ZZ'$-box diagrams to the observable scattering asymmetry for MOLLER experiment.
The only significant contribution to the Born asymmetry  comes from the interference terms from the $Z'$ and photon diagrams.
The relative correction to the Born asymmetry coming from $Z'$-boson is additive, and is given by
\begin{eqnarray}
\delta^{Z'}_A = \frac{v^{Z'}a^{Z'}}{v^{Z}a^{Z}}\frac{m_Z^{2}}{m_{Z'}^{2}}.
\end{eqnarray}

According to \cite{JLab12}, the goal of MOLLER is to measure the PV asymmetry to a precision of $2\%$ (0.73 ppb).
With this uncertainty, and assuming the identical coupling constants for $Z$ and $Z'$, it should be possible to detect ANB with a mass 
up to $m_{Z'} = \sqrt{m_Z^{2}/0.02} \approx 7 m_Z$. The sensitivity of
MOLLER to $Z'$ increases if its parity-violating couplings are larger
than those of $Z^0$, making the measurements of PV complementary to
the direct searches at high energies. 

The one-loop diagrams including ANB give significantly smaller contributions. As an example, let us consider $ZZ'$-box. 
As before, we perform our calculations by both approximate ("by-hand") and exact (with FeynArts and FormCalc) methods, and get an excellent agreement. 
The expressions derived as a result of our approximate approach are presented below.

For the $ZZ'$-box contribution, the cross section can be expressed by the following short equation:
\begin{eqnarray}
\sigma^{ZZ'-{\rm box}}=
\frac{3\alpha^3}{s} 
L
\sum\limits_{k=\gamma,Z}
\biggl[ \lambda_-^{B'k}(D^{kt}u^2+D^{ku}t^2)
\nonumber \\
  -2 \lambda_+^{B'k}s^2(D^{kt}+D^{ku}) \biggr] + (Z \leftrightarrow Z'),
\label{zzp}
\end{eqnarray}
where
\begin{eqnarray}
L= \frac{1}{m_Z^2-m_{Z'}^2} \log\frac{m_{Z'}}{m_Z},
\label{L-master}
\end{eqnarray}
and the functions $\lambda_{\pm}^{B'k}$ are expressed through
\begin{eqnarray}
  v^{B'}=v^Zv^{Z'} + a^Za^{Z'},\
  a^{B'}=v^Za^{Z'} + v^{Z'}a^Z.
\end{eqnarray}
To obtain $L$, we calculate the master scalar integral: 
\begin{eqnarray}
\int\limits_0^1 z^2 dz 
\int\limits_0^1 x  dx 
\int\limits_0^1 
\frac{dy}{m_Z^2z(x-1)+(t-m_{Z'}^2)(1-z)}.
\end{eqnarray}

According to the methods described in Section~\ref{sec41}, we can now calculate 
the relative correction to the observable asymmetry from the ANB contribution (i.e. from $ZZ'$-box)
$\delta_A^{\rm ANB}$ as:
\begin{eqnarray}
\delta_A^{\rm ANB}=
\frac{6\alpha m_Z^2}{\pi} \
\frac{v^{B'}a^{B'}}{v^{Z}a^{Z}}  \ L.
\end{eqnarray}
This correction is additive and becomes less important
with increasing $m_{Z'}$. 
However, this suppression is not very dramatic due to the growing log in the numerator.
If we take the MOLLER kinematics and assume that $v^{Z'}=v^{Z},\ a^{Z'}=a^{Z}$, then for $r_m\equiv m_{Z'}/m_Z=1$ 
the correction is twice the contribution from $ZZ$-box: $\delta_A^{\rm ANB} \approx -0.0025465$. 
As $r_m$ grows, the correction decreases:
at $r_m = 2$ the correction is $\delta_A^{\rm ANB} \approx -0.0011768$, and for 
$r_m = 10$ the correction is $\delta_A^{\rm ANB} \approx -0.0001185$. At $r_m=20$, the correction becomes 
completely negligible: $\delta_A^{\rm ANB} \approx -0.0000382$.
However, the possible contributions of new-physics particles to the M{\o}ller scattering deserves further attention, 
and we intend to continue our work in this direction. 

One of the simplest supersymmetric SM extensions is the Minimal
Supersymmetric Standard Model (MSSM), and it gives a useful framework
for discussing SUSY phenomenology. 
For $e^-e^-$ scattering, MSSM contributions will arise at the one-loop order, and the large suppression of the SM weak 
charge makes the weak charge sensitive to the effects of new physics. 
According to \cite{Mus2003}, the loop corrections in the MSSM can be
as large as $\sim 4\%$ for the weak charge 
of the proton and $\sim 8\%$ for the weak charge of the electron, 
which is close to the current level of experimental and theoretical precision available for the low-energy 
studies. Obviously, before we can interpret these high-precision 
scattering experiments in terms of possible new physics, it is crucial to have the SM EWC under a very firm control.

\section{Conclusion}
\label{sec:conclusion}

In the presented work, we perform
detailed calculations of the complete one-loop set of electroweak
radiative corrections to the parity violating $e^-e^- \rightarrow e^-e^-
(\gamma)$ scattering asymmetry both at low and high  energies using the on-shell
renormalization conditions proposed in \cite{Hol90} (see also \cite{BSH86})
and the conditions suggested in \cite{Denner}. 
Although contributions from the self-energies and vertex diagrams
calculated with the two sets of renormalization conditions  differ significantly, our full gauge-invariant set still guarantees that the
total relative weak corrections are in excellent agreement for the two
methods of calculation.

Obviously, it is important to exercise caution when comparing separate contributions arising
from the different renormalization conditions unless these contributions form a gauge-invariant 
set (like boxes). 
Although this is a well-known fact in principle, it is still useful to demonstrate 
this in detail numerically for a specific example. 
We hope that our results illustrating the structure 
of relative weak corrections evaluated at different renormalization conditions will be of educational 
value to researchers staring work in this area.

In addition, 
we compare the asymptotic results obtained analytically, "by hand" (with HRC), with some approximations, 
and semi-automatically (with DRC), with no approximations required.  
As a result, we have a good agreement for the whole $0 < \sqrt{s} < 50$~GeV energy region. 
More specifically, for the kinematics relevant to the 11 GeV MOLLER experiment planned at JLab, our agreement 
within two approaches for the complete one-loop set of
electroweak radiative corrections  is better than 0.1\%.
We found no significant theoretical uncertainty coming from the largest possible source, 
the had\-ro\-nic contributions to the vacuum polarization. 
The dependence on other uncertain 
input parameters, like the mass of the Higgs boson, is extremely weak and well below 0.1\%.
We conclude that the excellent agreement we obtained between the results calculated "by hand" 
and semi-automatically serves as a good illustration of opportunities offered by FeynArts, FormCalc, 
LoopTools,  and FORM. 

Considering the large size of the obtained 
radiative effects, it is obvious that the careful procedure for taking into account 
radiative correction is essential. 
Our plans include the construction of a Monte Carlo generator
for the simulation of radiative events within M{\o}ller
scattering to make our work directly useful to the experiment. 
Since we are now assured of the reliability of our calculations, we plan to base this Monte Carlo on the maximum-precision 
results from our semi-automatic approach.

Although making sure that the results obtained by two different approaches 
using two renormalization conditions are identical assures us
that our NLO EWC calculations are error-free, it does not address the question 
of the size of NNLO corrections. 
The two-loop corrections are beyond the scope of this work, but we plan to address them in the future. 
One way to find some indication of
the size of higher-order contributions is to compare physical
observables computed to the same order in perturbation theory in 
different renormalization schemes. 
Our calculations in the on-shell and CDR schemes show that while the
NLO terms differ by about 11\%, the PV asymmetries differ by about
3\%. At the level of precision of the future experiments such as
MOLLER,  higher-order corrections become important. 

To see if the two complimentary approaches we successfully used for the SM calculations can be applied in the NP domain, 
we expanded FeynArts, FormCalc, LoopTools, and FORM to include an additional neutral boson ($Z'$), 
calculated the relevant correction, and then obtained the same result by hand. Possible other contributions of new-physics particles to the M{\o}ller asymmetry still need 
to be investigated, and many of them can be included into the program packages mentioned above.

We believe that the future experiments at JLab and the ILC will mandate evaluation of the EWC beyond one loop.
Once all the SM corrections are under control, it is worth considering NLO corrections 
including new-physics particles, starting with the Minimal Super Symmetric Model (MSSM). 
The most straightforward way to address these corrections is by employing the CDR
scheme \cite{cdr2}, \cite{cdr1} because the CDR approach can be easily expanded to MSSM. 
However, whether the CDR scheme will be applicable 
in evaluating the EWC at the NNLO level  is still an open question. 
Our preliminary plan is to address the NNLO EWC with the on-shell scheme first, and 
if the effect is significant, stay with the same scheme for calculating contributions coming from the  
new-physics particles. 
The simple example of the $ZZ'$-box we consider in Section~\ref{sec5}
 is gauge-invariant and is 
thus not affected by the choice of renormalization, but we have to be careful when choosing 
the scheme for our future work. 
Any suggestions from the community regarding the best approach to this task would be greatly appreciated.

\section{ACKNOWLEDGMENTS}

We are grateful to T. Hahn, K. Kumar and E. Kuraev for stimulating discussions. 
A. A. and S. B. thank the Theory Center at JLab for hospitality in 2009 when this project was inspired. 
A. I. and V. Z. thank the Acadia and Memorial Universities for hospitality in 2010 and 2011. 
This work was supported by the Natural Sciences and Engineering Research Council of Canada
and the Belarussian State Program of Scientific Researches "Convergence".


\end{document}